%% file: 00_Main.tex
\documentclass[sigconf]{acmart}
\renewcommand\footnotetextcopyrightpermission[1]{} % removes footnote with conference information in first column

\usepackage{graphicx}
   \graphicspath{{./Figures/}}

\usepackage{amsmath}    
\usepackage{multirow}
\usepackage{svg}
\usepackage{color}
\usepackage{enumitem}

\DeclareGraphicsExtensions{.pdf,.jpeg,.png,.fig, .emf}
   \usepackage{subfigure}
    \usepackage{epstopdf}
    \usepackage{epsfig}
    \usepackage{subfigure}
    \usepackage{epstopdf}
    \usepackage{epsfig}

\usepackage{bm}

\usepackage{algorithm}
\usepackage{algorithmicx}
\usepackage{algpseudocode}

\copyrightyear{2024}
\acmYear{2024}
\setcopyright{acmlicensed}\acmConference[DAC '24]{61st ACM/IEEE Design Automation Conference}{June 23--27, 2024}{San Francisco, CA, USA}
\acmBooktitle{61st ACM/IEEE Design Automation Conference (DAC '24), June 23--27, 2024, San Francisco, CA, USA}
\acmDOI{10.1145/3649329.3655989}
\acmISBN{979-8-4007-0601-1/24/06}

\begin{document}
\title{\vspace{-10mm} 
Device-Algorithm Co-Design of \\ Ferroelectric Compute-in-Memory In-Situ Annealer  for \\ Combinatorial Optimization Problems }

 \author{\small{}%Xunzhao Yin, \IEEEmembership{Member, IEEE,}
         Yu Qian$^1$, 
         Xianmin Huang$^1$, %\IEEEmembership{Member IEEE,}
         Ranran Wang$^1$, Zeyu Yang$^1$, 
         Min Zhou$^1$, 
         Thomas K{\"a}mpfe$^{2,3}$, 
         %Kai Ni$^4$, %\IEEEmembership{Member IEEE,}
%         \\
         % Alptekin Vardar$^3$,
        Cheng Zhuo$^{1,4*}$
         and Xunzhao Yin$^{1,4*}$ %\IEEEmembership{Senior Member, IEEE}% <-this % stops a space
          \\$^1$%College of Information Science \& Electronic Engineering, 
          Zhejiang University, Hangzhou, China;
          $^2$Fraunhofer IPMS, Dresden, Germany; 
          $^3$TU Braunschweig, Braunschweig, Germany; 
          \\
          % $^4$University of Notre Dame, Notre Dame, USA;
          \vspace{-1ex}
          $^4$Key Laboratory of CS\&AUS of Zhejiang Province, Hangzhou, China;
          $^*$Corresponding author email: \{czhuo, xzyin1\}@zju.edu.cn
% \thanks{X. Yin, J. Cai, D. Gao and C. Zhuo are with the College
% of Information Science and Electronic Engineering, Zhejiang University, Hangzhou,
% China. E-mail: xzyin1@zju.edu.cn.}% <-this % stops a space
% \thanks{M. Imani is with the Department of Computer Science at UC Irvine, USA.}
% \thanks{K. Ni is with the Department of Electrical and Microelectronic Engineering, Rochester Institute of Technology, USA.}% <-this % stops a space
% \thanks{G.L. Zhang, B. Li and U. Schlichtmann are with the Chair of
% Electronic Design Automation with Technology University of Munich, Germany.}
% \thanks{C. Zhuo is the corresponding author. E-mail: czhuo@zju.edu.cn.}%
}

\renewcommand{\bibfont}{\small}
\let\oldbibliography\thebibliography
\renewcommand{\thebibliography}[1]{\oldbibliography{#1}
\setlength{\itemsep}{-0.5pt}} %Reducing spacing in the bibliography.

% % make the title area
% \maketitle

\begin{abstract}
\vspace{-1ex}
Combinatorial optimization problems (COPs) are crucial in many applications but are computationally demanding. 
%, including logistical and resource planning, network transportation, and chip and drug exploration. 
% To tackle COPs, many Ising annealers have been proposed, which typically involve direct transformation of COPs into Ising models and find the solution through iterative annealing process. 
% However, the computation of vector-matrix-vector (VMV) multiplication with the complexity of $O(n)$ and a complex exponential function are required in the annealing, resulting in substantially increased hardware costs. 
Traditional Ising annealers address COPs by directly converting them into Ising models (known as direct-E transformation) and solving them through iterative annealing. 
However, these approaches require vector-matrix-vector (VMV) multiplications with a complexity of $O(n^2)$ for Ising  energy computation and complex exponential annealing factor calculations during annealing process, thus significantly increasing hardware costs. 
% %a significant limitation arises in D-QUBO transformation 
% when addressing COPs with inequality constraints, this D-QUBO approach introduces numerous extra auxiliary variables, resulting in a substantially larger search space, increased hardware costs, and reduced solving efficiency.
In this work, we propose a  ferroelectric compute-in-memory (CiM) in-situ annealer to overcome aforementioned challenges. 
The proposed device-algorithm co-design framework consists of 
(i) a novel transformation method (first to our known) that converts COPs into an innovative incremental-E form, which reduces the complexity of VMV multiplication from $O(n^2)$ to $O(n)$, and approximates exponential annealing factor with a much simplified fractional form; 
(ii) 
%For the first time, we present an 
a double gate ferroelectric FET (DG FeFET)-based CiM crossbar that efficiently computes the in-situ incremental-E form by leveraging the unique structure of DG FeFETs; 
(iii) %When feasible solutions are detected,
a  CiM  annealer that approaches the solutions of COPs via iterative incremental-E computations within a tunable back gate-based in-situ annealing flow. 
% simulated annealing process.
% a CiM-based simulated annealing flow that 
% a FeFET-based CiM annealer that is capable of approaching global solutions of COPs via iterative QUBO computations within a 
% simulated annealing process.
%as a global acceleration mechanism.
%Evaluation results indicate that HyCiM significantly reduces the search space ($2^{200} \text{ to } 2^{2636}$) compared to the conventional D-QUBO approach, leading to substantial savings in hardware area overhead (90\% to 100\%). 
Evaluation results show that our proposed  CiM annealer significantly reduces hardware overhead, reducing energy consumption by 1503/1716$\times$ and time cost by 8.08/8.15$\times$ in solving 3000-node Max-Cut problems compared to  two state-of-the-art annealers. 
It also exhibits high solving efficiency, achieving a remarkable average success rate of 98\%, whereas other annealers show only 50\% given the same iteration counts.
% It also demonstrates high solving efficiency, successfully solving all COPs. 
% The evaluation results show that HyCiM drastically narrows down the search space, eliminating $2^{100} \text{ to } 2^{2536}$ infeasible input configurations compared to the conventional D-QUBO approach.
% Consequently, the narrowed search space, reduced to $2^{100}$ feasible input configurations, leads to a substantial hardware area overhead reduction, ranging from 88.06\% to 99.96\%.
% Additionally, 
% %HyCiM consistently ,
% %when faced with different initial states. Notably, 
% %even with the worst initial inputs, 
% %HyCiM consistently exhibits a high solving efficiency, achieving a remarkable success rate of 81\% to 100\% in solving COPs, whereas D-QUBO implementation shows only  1\%.
% HyCiM consistently exhibits a high solving efficiency, achieving a remarkable average success rate of 98.54\%, whereas D-QUBO implementatoin shows only 10.75\%.
\end{abstract}
%\vspace{-2ex}
% \begin{IEEEkeywords}
% Content addressable memory, associative search, nonvolatile memory, energy-aware scheme, FeFET.
% \end{IEEEkeywords}
% % no keywords

\maketitle
\pagestyle{empty}
%\vspace{-3mm}

% \fi
%\IEEEpeerreviewmaketitle

%\vspace{-3ex}
\input{01_Introduction}
\vspace{-2ex}
\input{02_Background}

\vspace{-2ex}

\input{03_Arch}

\vspace{-2ex}
\input{04_Evaluation}

\vspace{-2ex}
\input{05_Conclusion}

\section*{Acknowledgment}
%\vspace{-0ex}
%\begin{acks}
\vspace{-1ex}
This work was partially supported by NSFC (624B2126, W2412034, 92164203) and SGC Cooperation Project (Grant No. M-0612).
%\end{acks}
\vspace{-1.5ex}


%\vspace{-0.5ex}
% \section*{Acknowledgment}

% %The authors would like to thank...
% %\vspace{-0ex}
% %\begin{acks}
% \vspace{-1ex}
% This work was partially supported by NSFC (62104213, 92164203) and SGC Cooperation Project (Grant No. M-0612).
% %\end{acks}
% \vspace{-1.5ex}



%\bibliographystyle{IEEEtran}

%\bibliographystyle{ieeetr}
\newpage
\bibliographystyle{IEEEtran}
\bibliography{bib}

\end{document}

%% file: 01_Introduction.tex
\vspace{-2ex}
\section{Introduction}
\label{sec:intro}
\vspace{-1ex}

%Combinatorial optimization problems (COPs) find application across a wide array of domains such as logistics, resource allocation, communication network design, finance, drug discovery, and transportation systems, among others \cite{yu2013industrial, paschos2014applications, naseri2020application, barahona1988application}. Many of these problems are categorized as non-deterministic polynomial-time hard (NP-hard), signifying their position among the most intricate computational undertakings within the NP realm.

Combinatorial optimization problems (COPs) have widespread applications across various fields, including logistics, resource allocation, communication network design, and transportation systems \cite{yu2013industrial, paschos2014applications, naseri2020application, barahona1988application}, and
many of these problems are categorized as non-deterministic polynomial-time hard (NP-hard), 
reflecting their computational complexity.
%Addressing COPs through digital computers relying on the von Neumann architecture presents challenges due to the exponential surge in computational requirements and latency expectations \cite{markov2014limits, markov2013know, greenlaw1995limits}. 
%A notable recent advancement involves the realization that a significant array of COPs, including Max-Cut, graph coloring, knapsack problems, etc, can be effectively reformulated into Ising model or quadratic unconstrained binary optimization (QUBO) form —both of which are equivalent representations (See Sec. \ref{subsec:QUBO}) \cite{mohseni2022ising, date2021qubo, quintero2021characterizing}. 
%A recent notable advancement is the recognition that
A wide range of COPs, including Max-Cut, graph coloring, and knapsack problems, etc., can be effectively reformulated into Ising models or Quadratic Unconstrained Binary Optimization (QUBO) forms, which can be addressed by Ising machines and QUBO solvers 
%\cite{yin2024ferroelectric, mohseni2022ising, quintero2021characterizing}. 
\cite{jiang2023efficient, yue2024scalable, yin2024ferroelectric, mohseni2022ising}. 
%These representations are equivalent (refer to Sec. \ref{subsec:QUBO})
%Therefore, the remainder of this paper assumes that the Ising machine is capable of solving both Ising models and QUBO forms.
%
%These Ising machines and QUBO solvers, including digital ASIC annealers\cite{yamamoto20207, katsuki2022fast, onizawa2023local,takemoto20214} and dynamical system Ising machines \cite{moy20221, ahmed2021probabilistic}, etc., are specifically designed to harness nature's computational capability, thus can solve COPs much more quickly and efficiently than von Neumann hardware.
%
For example, digital ASIC annealers
%\cite{yamamoto20207, katsuki2022fast, onizawa2023local,takemoto20214} 
%\cite{yamamoto20207, katsuki2022fast,takemoto20214} 
\cite{katsuki2022fast,takemoto20214} 
implement different annealing algorithms within digital circuits, and dynamical system Ising machines \cite{moy20221, ahmed2021probabilistic} utilize the intrinsic system dynamics and their tendency to settle at ground state to solve COPs.
%These solvers are specifically designed to harness nature's computational capability, thus can solve COPs much more quickly and efficiently than von Neumann hardware.
These solvers address COPs more quickly and efficiently than von Neumann hardware, yet  still suffer from significant hardware overhead and limited robustness.
%
%and quantum Ising machines\cite{albash2018demonstration, denchev2016computational, boixo2016computational} 
%have been proposed to address Ising models or QUBO forms.
%This strategic conversion paves the way for their integration onto a computing-in-memory (CiM) crossbar-based system, often referred to as a QUBO solver, resulting in a remarkable acceleration of the pursuit for optimal solutions \cite{cai2020power, hong2021memory, taoka2021simulated}.

\begin{figure}[!t]
%\vspace{-3ex}
  \centering
  \includegraphics[width=1\columnwidth]{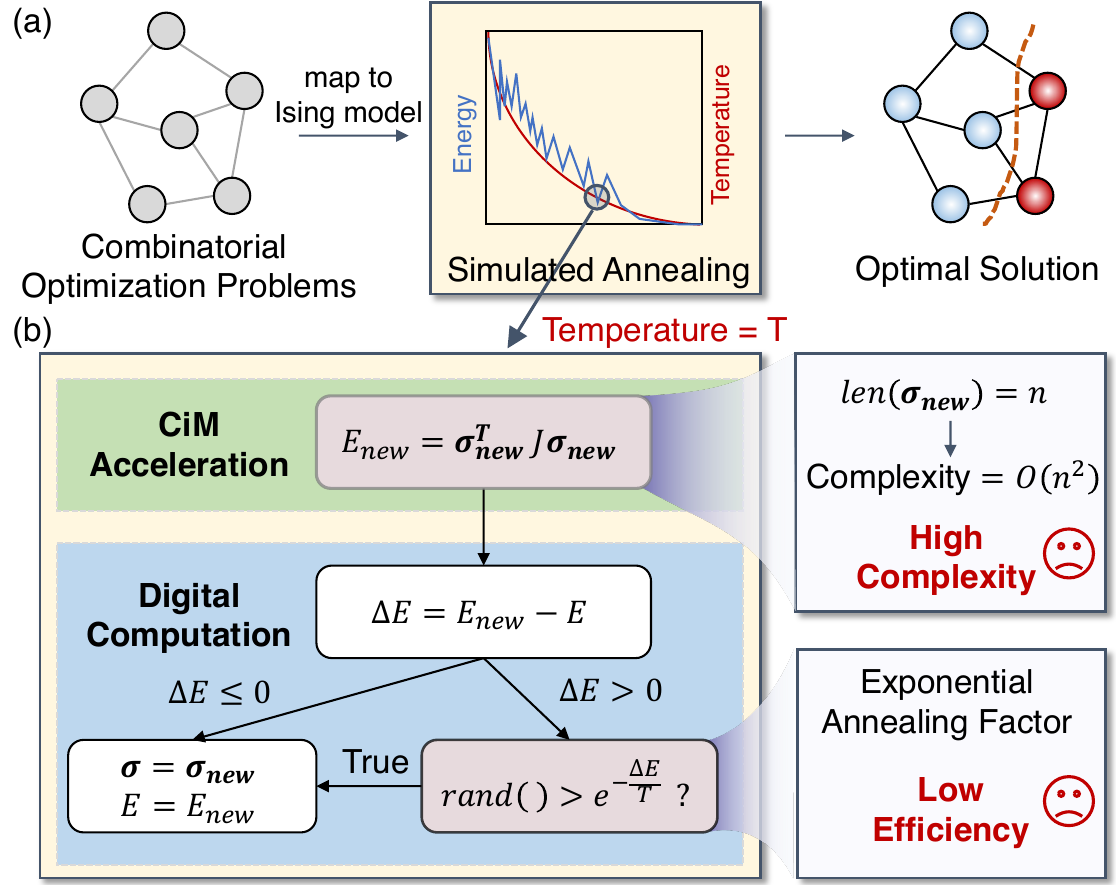}
  \vspace{-5ex}
  \caption{ CiM annealer solves COPs. 
  {\bf (a)} General flow based on simulated annealing algorithm; 
  {\bf (b)} Current solvers have high complexity in CiM acceleration part, and low efficiency in digital computation part.}
  \label{fig: motivation}
  \vspace{-4.5ex}
\end{figure}

Recently, numerous  compute-in-memory (CiM) frameworks based on non-volatile memories (NVMs) such as ferroelectric field effect transistor (FeFET) and resistive random access memory (ReRAM), have been proposed to further improve the hardware performance in solving COPs, known as CiM annealers \cite{taoka2021simulated, yin2024ferroelectric, qian2024c, qian2024hycim, qian2025ferroelectric}. 
Compared to the aforementioned solvers, CiM annealers utilize compact and highly parallel  NVM-based crossbar arrays to offer lower hardware costs and higher robustness.
As Fig. \ref{fig: motivation}(a) shows, 
%These CiM-based Ising machines utilize crossbar arrays to accelerate key operations, i.e., vector-matrix or vector-matrix-vector multiplications, and employ simulated annealing (SA) algorithms to approach optimal solutions.
after COPs are converted to a direct Ising model, 
CiM annealers compute the Ising energy within each simulated annealing iteration, progressively approaching optimal solutions as the temperature $T$ decreases.
As illustrated in Fig. \ref{fig: motivation}(b), at a given $T$, the annealer %updates the solution by leveraging CiM acceleration and digital computation. 
computes the new Ising energy through a CiM accelerated vector-matrix-vector (VMV) multiplication, and calculates the incremental energy difference $\Delta E$ in the digital circuits. 
Depending on $\Delta E$,
%If the value is non-positive, current solution is accepted and updated; otherwise 
an exponential annealing factor $e^{-\frac{\Delta E}{T}}$ is computed to decide whether to accept the new solution. Within the annealing process, CiM acceleration and digital computation are performed iteratively.
As can be seen, the CiM part involves a complex VMV multiplication with a computational complexity of $O(n^2)$, leading to large energy consumption and increased time cost \cite{kim2024scaling}.
Additionally, the digital computation part  requires calculating $e^{-\frac{\Delta E}{T}}$, which demands significant hardware resources, further reducing the computational efficiency \cite{hussain2022area}.

To address aforementioned challenges, we introduce 
%HyCiM, a hybrid computing-in-memory (CiM) QUBO solver framework that efficiently solves general COPs while minimizing hardware overhead and optimizing solving efficiency.
% an efficient  CiM in-situ annealer that significantly reduces the complexity of CiM acceleration and eliminates the need for complex computations in the digital domain.
a device-algorithm co-design approach of ferroelectric CiM in-situ annealer that not only significantly reduces  the complexity of direct Ising energy and exponential factor calculations, but also  enhances the  efficiency by leveraging the unique characteristics of 
%combines a novel Ising model annealing algorithm with an unique 
a double gate ferroelectric FET (DG FeFET)-based CiM crossbar for in-situ annealing. 
% to implement a novel Ising model transformation method and an in-situ annealing algorithm.  
%Unlike prior CiM annealers, the proposed annealer transforms the core computations of COP-converted direct Ising model into an incremental-E formula
%The  four-terminal structure and threshold voltage $V_{th}$-tunable back gate of the DG FeFET are utilized to efficiently support the costly . 
%These features enable CiM-based computation of a simplified objective function, following the proposed incremental-E transformation, and the implementation of an efficient in-situ annealing algorithm.
The key contributions of this work are summarized as follows:
\begin{itemize}[leftmargin=*]
    \item {We propose an incremental-E transformation method that converts a COP into an efficient incremental-E ($E_{inc}$) form. 
    %reducing the complexity of Ising energy computation from $O(n)$ to $O(1)$.
    The $E_{inc}$ form  eliminates redundant computations within the VMV multiplications, 
    and approximates complex exponential annealing factor with simpler fractional   annealing factor. 
    As a result,  the complexity of $E_{inc}$ computation is reduced from $O(n^2)$ to $O(n)$, significantly alleviating the associated hardware overhead.
    %The approach calculates the objective function with a constant time complexity of $O(1)$, leading to substantial savings in both time and energy consumption. 
    } 
    \item {
    We propose to realize a single DG FeFET-based four-input multiplication, as the device naturally supports the operator by applying three inputs to the double gates and drain, and storing an input as threshold voltage ($V_{TH}$) state, respectively.
    %DG FeFET, as the scalar inputs can be applied to the double gates, drain 
    Based on the unique characteristic of devices,
    we propose a DG FeFET-based  CiM crossbar array
    %that supports four-input multiplication by leveraging the double gate structure of DG FeFETs.
    %By mapping the spin vectors and coupling matrix elements  VMV multiplication and scalar fractional annealing factor to 
    that computes the entire $E_{inc}$ form, implementing the in-situ product of the VMV multiplication result and fractional annealing factor  within a single array.
    %by utilizing the double gate structure of DG FeFETs.
    %By utilizing the double gate structure of DG FeFETs, a four-input multiplication can be performe such  one-shot computations are realized based on the de 
    %Specifically, the DG FeFET's back gate (BG) is utilized to mimic the annealing temperature decay due to its threshold voltage tunability, thus realizing an in-situ annealing iteration by the CiM array.
    %fractional annealing factor is incorporated into the CiM acceleration by 
    } 
    \item {Based on the proposed transformation method and the unique CiM design, 
    %we present the CiM in-situ annealer, where  the DG FeFET's  back gate (BG) is utilized to mimic the annealing temperature decay  due to its threshold voltage tunability. 
    we propose a CiM in-situ annealer, where  the DG FeFET's  back gate (BG) is utilized to mimic the  temperature decay within the annealing flow due to its $V_{TH}$ tunability. 
    %that incorporates the annealing factor calculation and temperature decay directly into the CiM part, leaving only the solution update in digital part, and  approaching optimal solutions of COPs.
    %Compared with prior CiM annealers, 
    %our proposed device-algorithm co-design approach realizes an
    A tunable BG-based in-situ annealing flow is employed,  performing all the analog computations within the CiM array, and leaving only the solution update in digital part. 
    %benefits from the  costly temperature decay within the analog CiM crossbar, and leaves only the solution update in digital part, and  approaching optimal solutions of COPs.
    } 
    \item {The proposed CiM annealer demonstrates significant reductions in both time and energy consumption while achieving high accuracy in solving COPs. 
    In solving 30 Max-Cut problems with varying node numbers  from 800 to 3000, it offers 410$\times$ to 1706$\times$ energy reductions and 7.98$\times$ to 8.15$\times$  time cost reductions compared to other two state-of-the-art annealers. 
    It also achieves a 98\% success rate in average, while other annealers reach only 50\%, given the same iteration counts. 
    } 
\end{itemize}

%% file: 02_Background.tex
%\vspace{-0.5ex}
\section{Background}
\label{sec:background}
\vspace{-1ex}
In this section, we first introduce the Ising models and  related works, then review CiM and DG FeFET basics.

\vspace{-1ex}
%\subsection{QUBO Basics and COPs converted to QUBO}
\subsection{Ising models and existing solvers}
\label{subsec:Ising model}
\vspace{-1ex}
% \cmt{// Ising models --> SA flow --> other works --> current CiM works: flow improvement (MESA), operation improvement (compression) --> problems: complexity and e function}

%Ising models have emerged as a potent framework for adeptly modeling and resolving a diverse spectrum of COPs \cite{mohseni2022ising}. In this framework, problem variables are elegantly depicted as spins, while the interrelationships or constraints among variables manifest as spin couplings.
%The problem's objective function seamlessly translates into the Hamiltonian energy function of the Ising model. Consequently, problem solving involves identifying the spin configuration that minimizes the corresponding Ising Hamiltonian $H_{\mathrm{P}}$, which can be articulated as follows:
Ising models provide a powerful framework for modeling and solving COPs \cite{mohseni2022ising}, where problem variables are represented as spins with binary states, and constraints are captured through spin couplings. 
The objective function of the problem is formulated as the Hamiltonian energy function of the Ising model. 
Solving the problem involves finding the spin configuration that minimizes the corresponding Ising Hamiltonian $H_{\mathrm{P}}$:
\vspace{-1ex}
\begin{equation}
\vspace{-1ex}
\label{equ:Ising model}
    \small \min H_{\mathrm{P}}=\sum_{i,j=1}^{n}J_{ij}\sigma_{i}\sigma_{j}+\sum_{i=1}^{n}h_{i}\sigma_{i}
%\vspace{-0.5ex}
\end{equation}
where $n$ denotes the number of spins and $\sigma_i \in\pm1$ represents the state of spin $i$. $J_{ij}$ and $h_i$ stand for the coupling between spin $i$ and $j$ and the self-coupling of spin $i$, respectively.
%, governing the spins' behavior.
Eq. \eqref{equ:Ising model} can be expressed in an alternative general form by setting $J_{ii}=h_i$:
\vspace{-1ex}
\begin{equation}
\vspace{-1ex}
\label{equ: general Ising model}
    \small \min E = \bm{\sigma^T} J \bm{\sigma}
%\vspace{-0.5ex}
\end{equation}
where $\bm{\sigma}$ presents the spin vector.
The Ising model is equivalent to the QUBO form \cite{glover2018tutorial}  through a simple variable transformation $\sigma_i = 1 - 2x_i$, $x_i \in \{0,1\}$. Thus, the Ising machines and QUBO solvers can be regarded as equivalent COP solving designs.

Numerous Ising machines %and QUBO solvers 
have been developed to address COPs. 
Digital ASIC annealers, which implement various annealing algorithms to solve Ising models \cite{katsuki2022fast, takemoto20214}, often suffer from high energy consumption and time costs due to the exponential growth of explicit computations in each annealing iteration. 
Dynamic system Ising machines, on the other hand, exploit the natural dynamics of physical systems to explore optimal solutions \cite{mallick2023cmos, afoakwa2021brim}, but they are highly sensitive to spin coupling implementations. Even slight deviations in coupling strength can hinder convergence \cite{ahmed2021probabilistic, mallick2023cmos}.

To further enhance the robustness and hardware efficiency of solving COPs, 
a novel framework known as
the CiM annealer has been proposed recently \cite{yin2024ferroelectric, qian2024hycim, qian2024c}. 
% CiM annealer accelerates the key operation (Eq. \eqref{equ: general Ising model}) by utilizing the CiM architecture.
Fig. \ref{fig: motivation}(a) depicts the general flow of the CiM annealer. 
The COPs, e.g., Max-Cut problems, are formulated as Ising models.
The Ising energy $E$ is progressively minimized as  temperature decreases, following the simulated annealing (SA) algorithm. 
%In each iteration, i.e., a given temperature point, 
At each temperature, the value of $E_{new}$ based on updated spin states $\bm{\sigma_{new}}$ is  computed using the CiM architecture, followed by peripheral digital computations. 
% Several CiM annealers have been proposed these years.
A FeFET-based CiM annealer in \cite{yin2024ferroelectric} 
provides a lossless compression method to reduce the cost of CiM architecture and enhances the SA algorithm into Multi-Epoch SA (MESA) for improved performance. 
The HyCiM framework in \cite{qian2024hycim} addresses general COPs with inequality constraints by employing an inequality filter CiM structure. 
%by employing an 'inequality filter' CiM architecture to handle these constraints. 
%In \cite{qian2024c}, a
A C-Nash CiM framework is proposed to solve COPs with the MAX function embedded in the objective function \cite{qian2024c}.
While these CiM annealers leverage the CiM  to 
reduce energy and time costs and enhance SA algorithm efficiency, 
%However, 
they still face challenges on hardware overhead:

% However, the operations within each iteration remain energy- and time-intensive, limiting the overall efficiency of the CiM annealer. 
As shown in Fig. \ref{fig: motivation}(b), the core VMV multiplication operation  %accelerated by the CiM crossbar 
within each annealing iteration 
%However, the operation 
involves $n^2$ product terms given $n$ spins, exhibiting a high complexity of $O(n^2)$.
% \vspace{-2ex}
% \begin{equation}
% \vspace{-1ex}
% \label{equ: sum of all columns}
%     \small E \ = \bm{\sigma_{new}^T} J \bm{\sigma_{new}}
%       \ = \sum_{i=1}^n \bm{\sigma_{new}^T} J \sigma_{new,i}
%       \ = \sum_{i=1}^n C_i
% %\vspace{-0.5ex}
% \end{equation}
% where the values of $C_i$s are usually identified sequentially by multiplexed analog-to-digital converters (ADCs).
% % The input configuration $\bm{\sigma_{new}}$ has a dimension equal to the spin number $n$, making the size of the CiM crossbar proportional to $n^2$, considering quantization. 
% Consequently, the hardware overhead including both energy and time costs
% %the energy and time consumption in both the CiM crossbar and peripheral circuits, 
% scales with a high complexity of $O(n)$. 
% As the problem size increases, e.g., $n=3000$, the decreasing of efficiency is inevitable even only $1$ or $2$ spins are flipped in each iteration. 
Additionally, as shown in Fig. \ref{fig: motivation}(b), a complex exponential annealing factor calculation is required when $\Delta E>0$,  further reducing the efficiency of the CiM annealer.
This has inspired us to propose an incremental-E transformation  that reduces the complexity of VMV multiplication from $O(n^2)$ to $O(n)$, and eliminates the costly exponential function. 
Such incremental-E form is then implemented by a CiM array.
%, where the back gate of DG FeFET device is leveraged to proceed annealing iterations.
% To address this issue, we propose a CiM architecture which directly embeds the complex exponential function, thereby eliminating the associated computational overhead.

\begin{figure}[!t]
%\vspace{-3ex}
  \centering
  \includegraphics[width=1\columnwidth]{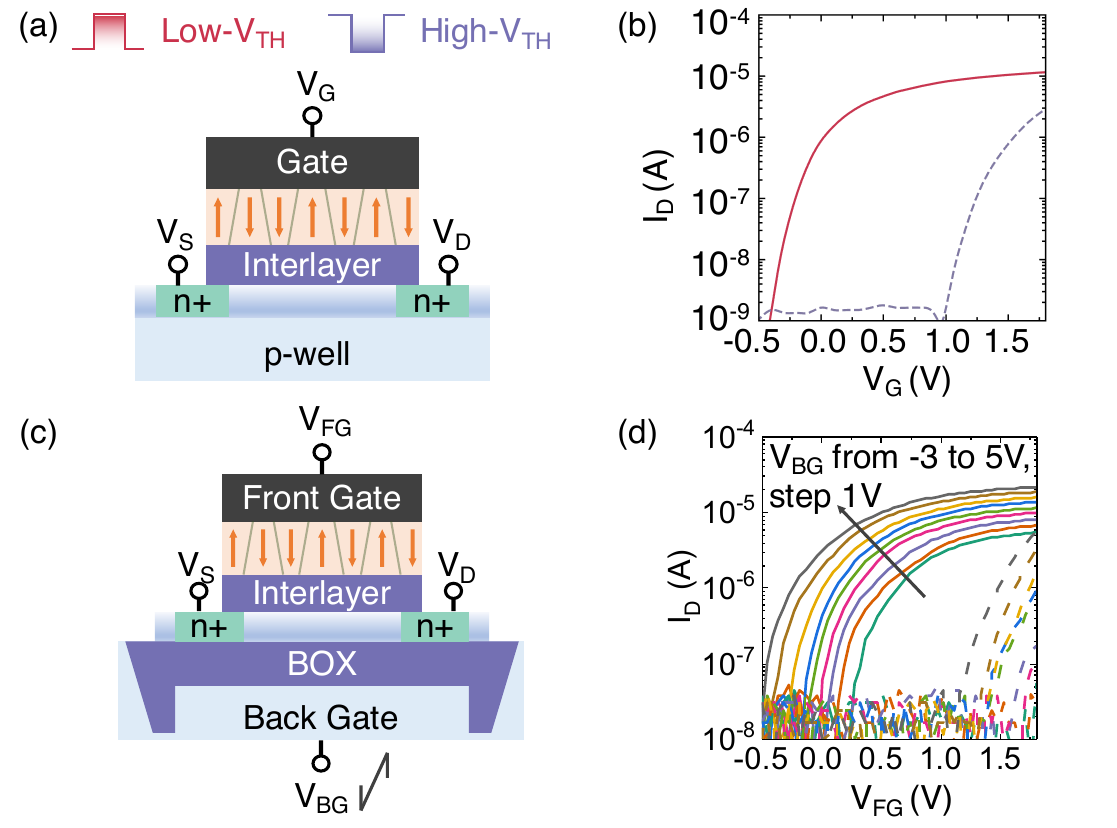}
  \vspace{-6ex}
  \caption{ 
  FeFET and DG FeFET. 
  (a) A FeFET device with three terminals; 
  (b) $I_D-V_G$ curves of a FeFET with programmed low/high $V_{TH}$.
  (c) A DG FeFET device with four terminals;
  (d) $I_D-V_G$ curves of the DG FeFET, where $V_{BG}$ can flexibly adjust its $V_{TH}$ after programming.
  }
  \label{fig:FeFET device}
  \vspace{-4ex}
\end{figure}

\vspace{-1ex}
\subsection{CiM Preliminaries and DG FeFET Basics}
\label{subsec:CiM Preliminary}
\vspace{-1ex}

Prior CiM designs employ NVMs, such as ReRAM, magnetic tunneling junction (MTJ) and FeFET to build crossbars that store  weight matrices, and perform parallel vector-matrix multiplication or VMV multiplication operations 
%for neural network accelerations 
\cite{zhuo2022design, qian2024enhancing, zhao2024convfifo, mondal2018situ, yin2024ferroelectric, yin2024homogeneous}.
% However, these NVMs still face challenges for enhancing energy and area efficiency due to the current-driven mechanism, low ON/OFF ratio and large driving transistors. 
% FeFET, which integrates HfO$_2$ as the ferroelectric dielectric within the gate stack of MOSFET, has  emerged as a promising candidate for embedded memory and CiM.
% Compared with  ReRAM \cite{salahuddin2018era} and MTJ \cite{zhuo2022design}, 
Among these devices, FeFET offers advantages in energy and area efficiency due to the  CMOS compatibility, voltage-driven mechanism, high ON/OFF ratio and three-terminal structure \cite{ni2019ferroelectric, xu2023challenges, cai2024scalable,  yin2023ultracompact, yin2024deep, hu2021memory, yin2022ferroelectric}.
When the gate is applied with different gate pulses as shown in Fig. \ref{fig:FeFET device}(a), %shows that how 
a FeFET can store  high or low $V_{TH}$ states, which are experimentally measured from the FeFET \cite{yin2024ferroelectric} as shown in Fig. \ref{fig:FeFET device}(b). %\textcolor{blue}{Put reference to the data here.}
% While this binary storage characteristic of FeFETs can be beneficial for CiM circuits such as crossbar \cite{qian2024c}, 
% the value of $V_{TH}$ remains fixed after programming unless the FeFET is erased and reprogrammed, 
% limiting its application to scenarios that require frequent adjustments of $V_{TH}$. 
Prior works exploit the intrinsic three-input multiplication operator of FeFETs for  VMV multiplication acceleration, thus enabling efficient CiM-based COP solvers \cite{yin2024ferroelectric, qian2024c, qian2024hycim}. 
That said, the potentials of ferroelectric devices are yet to be explored.
%Due to the binary storage characteristic of FeFETs, where $V_{TH}$ remains fixed after programming unless erased and reprogrammed, FeFETs are inherently unsuitable for high-performance applications requiring temporary adjustments to $V_{TH}$. 
%\textcolor{blue}{This sentence does not make sense.}

Recently, a novel technology known as DG FeFET \cite{jiang2022asymmetric} has been developed, which incorporates a non-ferroelectric (non-FE) dielectric in the BG. A notable example of such a device structure is fully-depleted silicon-on-insulator (FDSOI) FeFET, as shown in Fig. \ref{fig:FeFET device}(c), where the buried oxide (BOX) layer serves as the non-ferroelectric dielectric. 
Unlike the fixed $I_D-V_G$ curves of a conventional FeFET, the $I_D-V_G$ curves for  the low-$V_{TH}$/high-$V_{TH}$ state of a DG FeFET can be dynamically adjusted by applying different voltage on BG, as illustrated in Fig. \ref{fig:FeFET device}(d). 
This  $V_{TH}$ tuning does not change the ferroelectric states, but is rather due to the electric field exerted through the BG. 
This unique BG-based $V_{TH}$ tuning is used in this work to create a cell that efficiently operates with three input terminals (i.e., double gates and drain), enabling in-situ  annealing computations.
%enhancing the  computational efficiency and annealing flexibility  of CiM annealer.

%% file: 03_Arch.tex
\section{Proposed CiM Annealer}
\label{sec:design}
\vspace{-1ex}

%In this section, 
%we introduce HyCiM, a hybrid CiM QUBO solver framework for COPs with inequality constraints.
%building upon the motivation outlined in Sec. \ref{sec:background}.
Here we introduce our proposed device-algorithm co-design approach, including the incremental-E transformation, the  DG FeFET-based CiM  array and tunable BG-based in-situ annealing flow. 
%outlining its core components and operational flow. We 
%then discuss  core  components with operational flow.

\begin{figure}[!t]
%\vspace{-3ex}
  \centering
  \includegraphics[width=1\columnwidth]{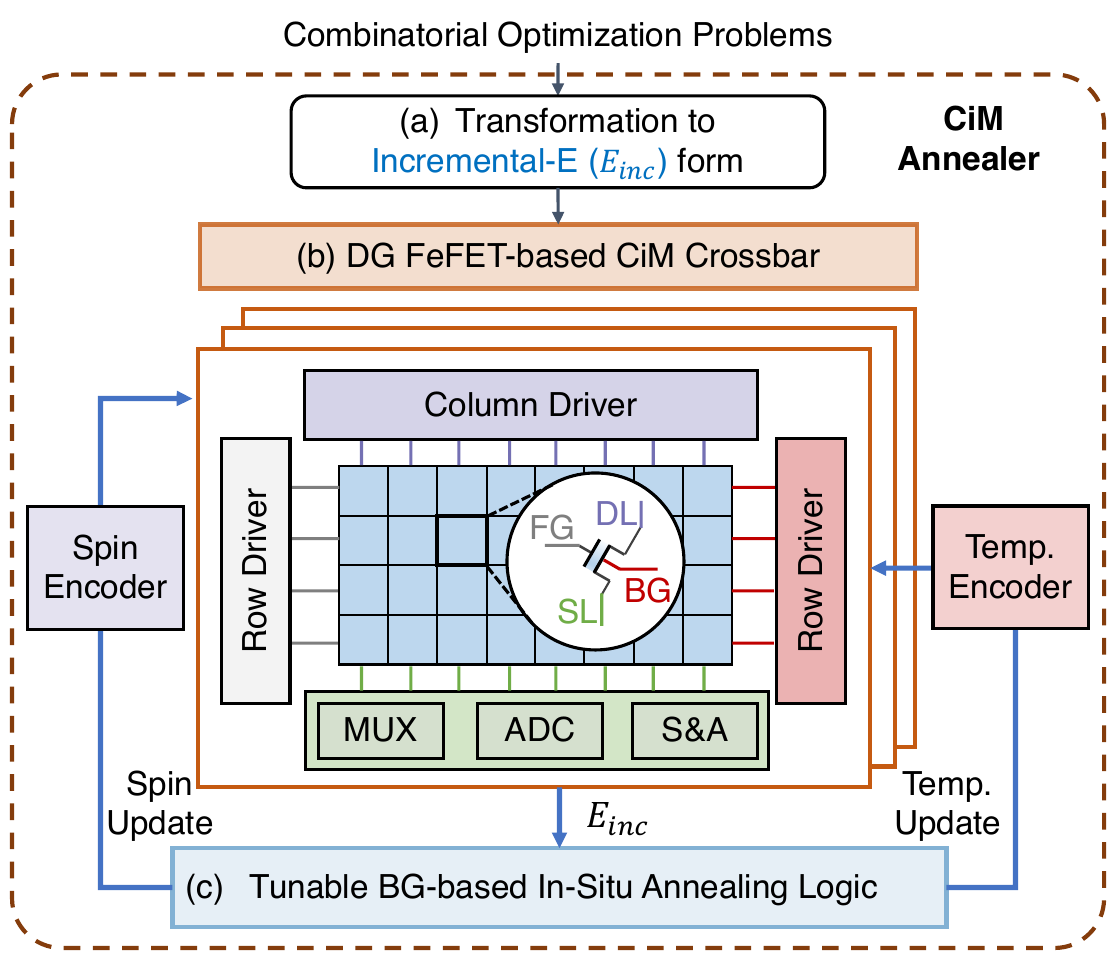}
  \vspace{-5ex}
  \caption{Overview of the proposed CiM annealer. 
  (a) Incremental-E ($E_{inc}$) transformation method; 
  (b) DG FeFET-based  crossbar for $E_{inc}$ computation; 
  (c) Tunable BG-based in-situ annealing  for approaching optimal solutions.
  }
  \label{fig:overview}
  \vspace{-4ex}
\end{figure}

\vspace{-1ex}
\subsection{Overview}
\label{subsec:overview}
\vspace{-1ex}

Fig. \ref{fig:overview} illustrates the overview of the proposed  CiM annealer. 
The COP is firstly transformed into an incremental-E ($E_{inc}$) form (Sec. \ref{subsec:transformation}), 
which is then mapped onto 
%dedicated hardware implementations, namely 
the proposed DG FeFET-based CiM crossbar for  computation (Sec. \ref{sec:crossbar}).
Finally, the tunable BG-based in-situ annealing flow is employed, approaching optimal solutions
%incorporated into the CiM computation within each iteration, achieving in-situ annealing  flow
(Sec. \ref{subsec: flow}).

During each annealing iteration, the tunable BG-based annealing flow updates the spin states based on the previous $E_{inc}$ value, transmitting  solely  the updated spins to the spin encoder. 
The updated temperature is relayed to the temperature encoder. 
% These two encoders then send the corresponding information to the DG FeFET-based crossbar for further computations.
The DG FeFET-based CiM array is then activated to compute the new $E_{inc}$, which is fed back to the annealing logic to initiate  next iteration.

\vspace{-1ex}
\subsection{Incremental-E Transformation}
\label{subsec:transformation}
\vspace{-1ex}

\begin{figure}[!t]
%\vspace{-3ex}
  \centering
  \includegraphics[width=1\columnwidth]{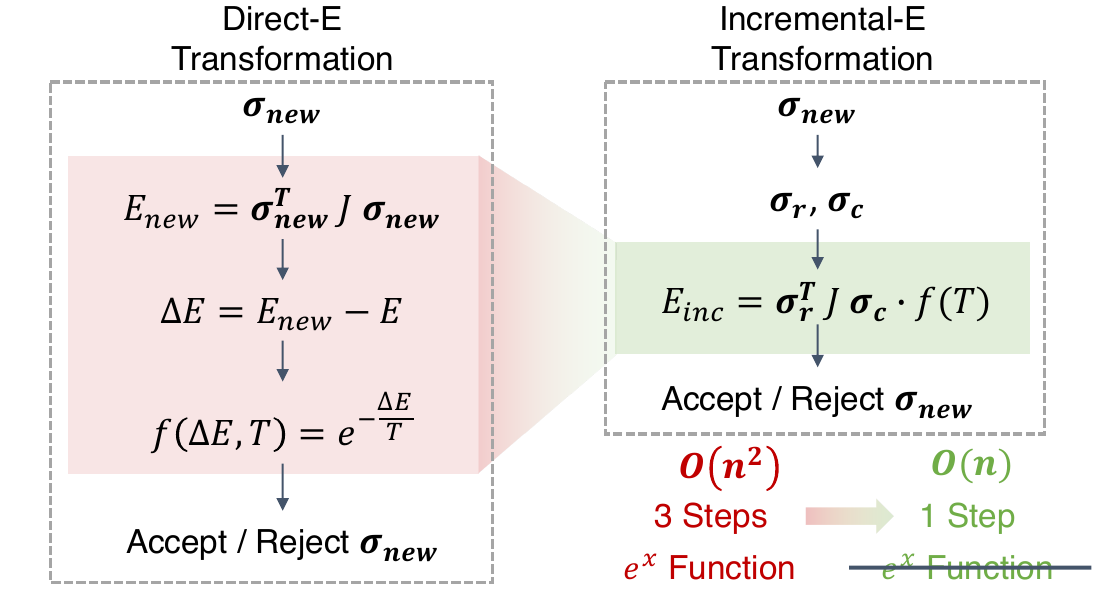}
  \vspace{-5ex}
  \caption{
  %Incremental-E transformation.
  % (a) Current direct-E transformation method requires mainly three steps within an iteration; 
  % (b) The proposed incremental-E  approach condenses  three steps into a single operation; 
  % (c) 
  Compared to direct-E, the incremental-E transformation reduces complexity from $O(n^2)$ to $O(n)$, condenses  three steps into a single operation,  
  %minimizes the steps, 
  and eliminates $e^x$ function. 
  }
  \label{fig: Incremental-E}
  \vspace{-4ex}
\end{figure}

% Note that within each iteration, two operations in current CiM aneealers are inevitable:
% As illustrated in Fig. \ref{fig: Incremental-E}(a), the direct-E transformation method in current CiM annealers requires three  operations on the critical path within an iteration:
% In current CiM annealers, the operations within an SA iteration rely on a direct-E transformation method, as illustrated in Fig. \ref{fig: Incremental-E}(a). 
% This method requires three complex operations within the critical path:
%\polish{
% As shown in Fig. \ref{fig: Incremental-E}, the proposed incremental-E method reduces the complexity from \(O(n)\) to \(O(1)\) compared to direct-E transformation as the number of flipped spins \(|F|\) is set to be constant. 
% This approach also merges the three critical computations into a single step, as 
% the resulted incremental-E $E_{inc}$ formulation can be efficiently implemented by  our proposed DG FeFET-based CiM crossbar. 
Fig. \ref{fig: Incremental-E} shows the proposed incremental-E transformation method, which 
replaces the costly VMV multiplication of  $E_{new}$ and $e^x$ function in the direct-E transformation 
%(Fig. \ref{fig: Incremental-E}(a)) 
with the  product of an incremental-based VMV multiplication and a fractional factor, thus reducing 
%(Fig. \ref{fig: Incremental-E}(b)). 
%Fig. \ref{fig: Incremental-E}(c) illustrates that 
%The proposed incremental-E form reduces 
the complexity from \(O(n^2)\) to \(O(n)\). 
Such transformation merges the three critical computations into a single step, and eliminates the need for $e^x$ calculation.

\begin{figure}[!t]
%\vspace{-3ex}
  \centering
  \includegraphics[width=1\columnwidth]{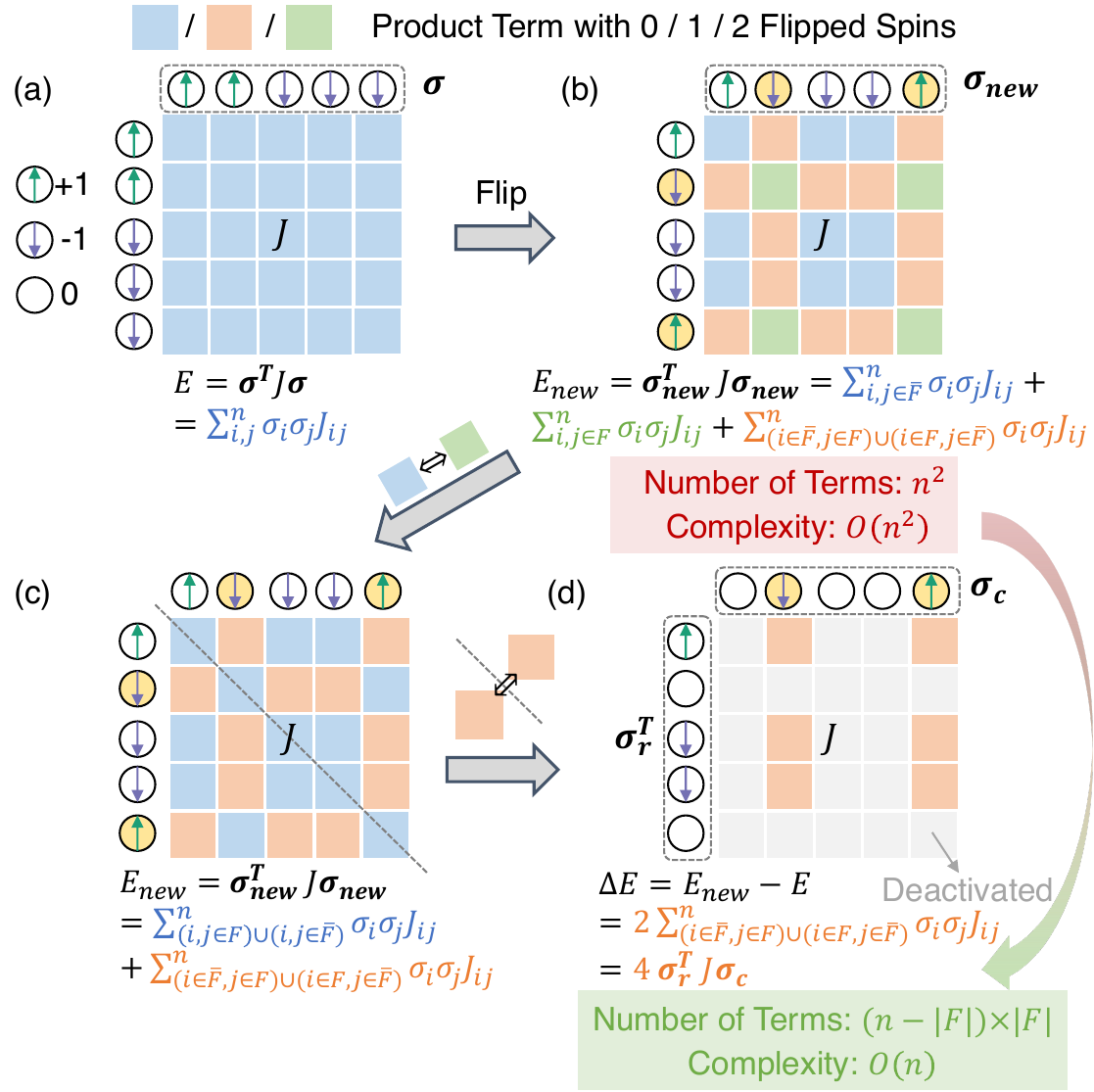}
  \vspace{-5ex}
  \caption{
  Complexity reduction of VMV multiplication in the incremental-E transformation. 
  % Incremental-E transformation illustration. 
  After (a) $\bm{\sigma}$ is flipped to (b) $\bm{\sigma_{new}}$, $E_{new}$ has a complexity of $O(n^2)$,  composing of three groups of product terms in different colors; 
  (c) The distinct terms between $E$ and $E_{new}$ in orange are   merged by diagonal symmetry, resulting in (d) $\Delta E$  with $O(n)$ complexity. 
  }
  \label{fig: compression}
  \vspace{-4ex}
\end{figure}

As shown in Fig. \ref{fig: motivation}(b), current annealers calculate $\Delta E$ within an iteration as follows: 
% Within an iteration, the old spin vector $\bm{\sigma_{old}}$ 
\vspace{-1ex}
\begin{equation}
\vspace{-1ex}
\label{equ: E new}
    \small E_{new} = \bm{\sigma_{new}^T} J \bm{\sigma_{new}}
\vspace{-1ex}
\end{equation}
\vspace{-1ex}
\begin{equation}
\vspace{-1.5ex}
\label{equ: E substraction}
    \small \Delta E = E_{new} - E
\vspace{1ex}
\end{equation}
% % \vspace{-1ex}
% \begin{equation}
% \vspace{-1.5ex}
% \label{equ: e^x}
%     \small f(\Delta E, T) = e^{-\frac{\Delta E}{T}}
% \vspace{1ex}
% \end{equation}
where $E$ and $E_{new}$ denote the current and new Ising energy, and $\bm{\sigma}$ and $\bm{\sigma_{new}}$ denote the current and new spin vectors, respectively. 
%Note that the VMV multiplication in Eq. \eqref{equ: E new} has a complexity of $O(n^2)$.
% and the exponential annealing factor in Eq. \eqref{equ: e^x} involves the costly exponential function $e^x$. 
% To optimize the direct-E approach, we propose an incremental-E transformation method, as illustrated in Fig. \ref{fig: Incremental-E}(b). 
% As $n$ increases, the efficiency is decreased.
% This has inspired us to propose a new transformation method to avoid 
% Fig. \ref{fig: Incremental-E}(a) illustrates the process in calculating $\Delta E$ in current CiM architectures, where the number of flipped spins in each iteration is fixed at $2$.
% As shown in Fig. \ref{fig: Incremental-E}(b), 
Fig. \ref{fig: compression} shows the process of complexity reduction in the proposed incremental-E transformation. 
Within an iteration, the current spin vector $\bm{\sigma}$ (Fig. \ref{fig: compression}(a)) is flipped to generate $\bm{\sigma_{new}}$ (Fig. \ref{fig: compression}(b)). 
The corresponding $E_{new}$ can be decomposed as:
\begin{equation}
\vspace{-1.5ex}
\label{equ: E new decomposition}
    \small E_{new} = \sum_{i,j \in \overline{F}}\sigma_i\sigma_jJ_{ij} 
            + \sum_{i,j \in F}\sigma_i\sigma_jJ_{ij}
            + \sum_{\substack{i \in F, j \in \overline{F}}} \sigma_i\sigma_j J_{ij}
            + \sum_{\substack{i \in \overline{F}, j \in F}} \sigma_i\sigma_j J_{ij}
\vspace{1ex}
\end{equation}
where each spin \( \sigma_i \in \{-1,1\} \), \( F \) denotes the index set 
%containing indices 
of flipped  spins, 
and \( \overline{F} \) denotes the rest set. 
The computation of $E_{new}$ entails $n^2$ product terms,  exhibiting a complexity of $O(n^2)$. 
These terms can be categorized into  three groups, i.e., $\sigma_i\sigma_jJ_{ij}$ with 0/1/2 flipped spins, as shown in Fig. \ref{fig: compression}(b), corresponding to the boxes in blue/orange/green   within the spin coupling matrix. 
%Specifically,  the blue/orange/green color represents a product term $\sigma_i\sigma_jJ_{ij}$ with 0/1/2 flipped spins. 
If spin $i$ and $j$ are both flipped (switch to the other binary state), the  term $\sigma_i\sigma_jJ_{ij}$ remains unchanged. Therefore, the green terms can be recolored to blue, as depicted in Fig. \ref{fig: compression}(c). 
% We can derive the following relationship: 
% \begin{equation}
% \vspace{-1.5ex}
% \label{equ: E derive 1}
%     \small \sum_{i,j \in \overline{F} F}\sigma_i\sigma_jJ_{ij} = \sum_{i,j \in {F}}\sigma_i\sigma_jJ_{ij}
% % \vspace{1ex}
% \end{equation}
$E_{new}$ can then be expressed as: 
\begin{equation}
\vspace{-1.5ex}
\label{equ: E derive 2}
    \small E_{new} = \sum_{(i,j \in F)\cup(i,j \in \overline{F})}\sigma_i\sigma_jJ_{ij} 
            + \sum_{(i\in \overline{F}, j\in F)\cup(i\in F, j\in \overline{F})} \sigma_i\sigma_j J_{ij}
\vspace{1ex}
\end{equation}
where 
%the distinct terms between $E_{old}$ and $E_{new}$, i.e., 
the latter term in orange corresponds to the distinct terms between $E$ and $E_{new}$.
%in Eq. \eqref{equ: E derive 2}, are denoted by orange boxes.  
These orange terms can be merged  along the diagonal of matrix  due to the symmetry. 
% Eq. \eqref{equ: E derive 1} shows that if spin $i$ and $j$ are both flipped  (switch to the other binary state), the  term $\sigma_i\sigma_jJ_{ij}$ remains unchanged, as flipping both spins is equivalent to no flip occurring for these spins. 
% Eq. \eqref{equ: E derive 2} holds for the same reason, demonstrating that the scenario where spin $i$ is flipped and $j$ is not flipped is equivalent to the scenario where spin $i$ is not flipped and $j$ is flipped, with the product term $\sigma_i\sigma_jJ_{ij}$ having the opposite sign compared to when both spins are flipped. 
% Thus, Eq. \eqref{equ: E new decomposition} simplifies to:
% \begin{equation}
% \vspace{-1.5ex}
% \label{equ: E new to old}
%     \small E_{new} = E_{old} - 2 \sum_{\substack{i \in F, j \notin F \cup i \notin F, j \in F}} \sigma_i \sigma_j J_{ij}
% \vspace{1ex}
% \end{equation}
% Consequently, \(\Delta E\) is expressed as:
% \begin{equation}
% \begin{aligned}
% \vspace{-1.5ex}
% \label{equ: delta E detail}
%     \small \Delta E = -2 \sum_{\substack{i \in F, j \notin F \cup i \notin F, j \in F}} \sigma_i \sigma_j J_{ij}
% \vspace{1ex}
% \end{aligned}
% \end{equation}
After the merging, we define a logical vector \(\bm{\sigma_f}\), where \(\sigma_{f,i} = 1\) indicates that spin \(i\) is flipped. 
We also define two vectors, \(\bm{\sigma_c}\) and \(\bm{\sigma_r}\), based on 
%the updated configuration 
\(\bm{\sigma_{new}}\):
\vspace{-1ex}
\begin{equation}
\label{equ: sigma column}
    \small \bm{\sigma_{c}} = \bm{\sigma_{new}} \circ \bm{\sigma_{f}}
\vspace{-1.5ex}
\end{equation}
\vspace{-2ex}
\begin{equation}
\vspace{-2ex}
\label{equ: sigma row}
    \small \bm{\sigma_{r}} = \bm{\sigma_{new}} \circ (1 - \bm{\sigma_{f}})
\vspace{1ex}
\end{equation}
where \(\circ\) denotes element-wise multiplication. 
The vector \(\bm{\sigma_c}\) maintains the values of the flipped elements in \(\bm{\sigma_{new}}\), setting all others to zero, 
while \(\bm{\sigma_r}\) preserves the unflipped spin values, 
%from \(\bm{\sigma_{new}}\), 
zeroing out the flipped elements.
As a result, $\Delta E$ can be expressed as an incremental-based VMV multiplication:
\vspace{-1ex}
\begin{equation}
\label{equ: sigma column sum}
    \small \Delta E = 4\bm{\sigma_{r}^T} J \bm{\sigma_{c}} 
\vspace{-1ex}
\end{equation}
where the number of  product terms is only $(n-|F|)\times|F|$, exhibiting a reduced complexity of $O(n)$ because \(|F|\) is set to be constant, as depicted in Fig. \ref{fig: compression}(d). 
%}

The exponential annealing factor in Fig. \ref{fig: motivation}(b) is approximated as 
\vspace{-1ex}
\begin{equation}
\label{equ: taylor}
    \small e^{-\frac{\Delta E}{T}} \approx 1 - \frac{\Delta E}{T} \propto - \frac{\Delta E}{T}
\vspace{-0.5ex}
\end{equation}
Therefore, after simplification, the incremental-E form is as follows:
\vspace{-2ex}
\begin{equation}
\label{equ: incremental-E}
    \small E_{inc} = \bm{\sigma_{r}^T}  J  \bm{\sigma_{c}} \cdot f(T) 
\vspace{-0.5ex}
\end{equation}
where \(f(T) = \frac{a}{bT + c} + d\), with parameters \(a\), \(b\), \(c\), and \(d\) properly chosen. 
Note that Eq. \eqref{equ: incremental-E} must satisfy 
%two constraints:
(i) $f(T) > 0$, as $e^{-\frac{\Delta E}{T}} > 0$, and (ii) $f(T)\propto T$. 
% As shown in Fig. \ref{fig: Incremental-E}, 
% As a result, the proposed incremental-E method reduces the complexity 
% from \(O(n^2)\) to \(O(n)\) compared to direct-E transformation as the number of flipped spins \(|F|\) is set to be constant. 
% This approach also merges the three critical computations into a single step, as 
% the resulted incremental-E $E_{inc}$ formulation can be efficiently implemented by  our proposed DG FeFET-based CiM crossbar. 
The resulted $E_{inc}$ formulation can be efficiently implemented by  our proposed DG FeFET-based CiM crossbar.

\vspace{-1ex}
\subsection{DG FeFET-based CiM Crossbar}
\label{sec:crossbar}
\vspace{-1ex}
% \cmt{
% 1. cell characteristics with temp
% 2. crossbar architecture
% 3. delta-E realization
% }

\begin{figure}[!t]
%\vspace{-3ex}
  \centering
  \includegraphics[width=1\columnwidth]{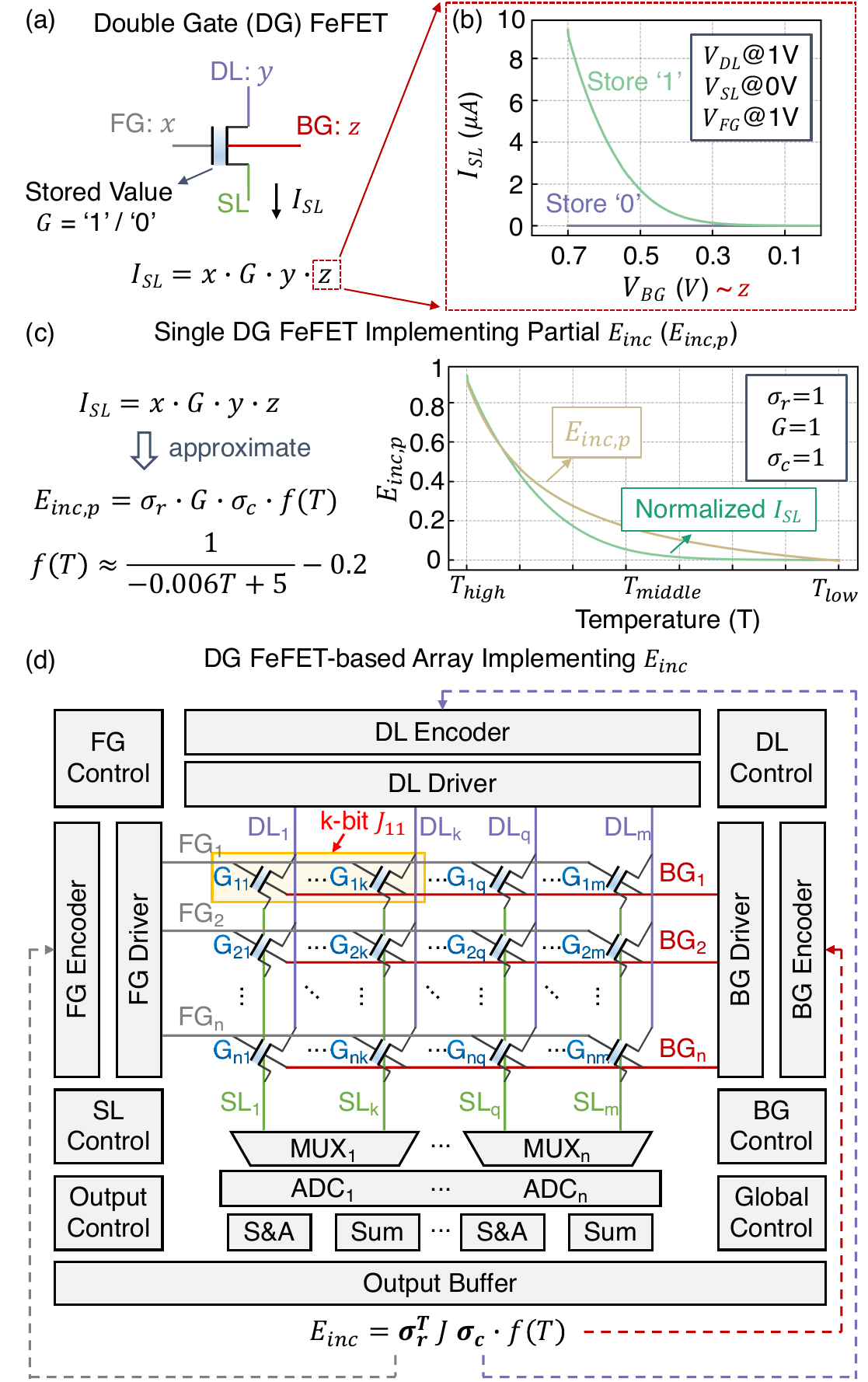}
  \vspace{-5ex}
  \caption{
  DG FeFET-based CiM crossbar array.
  (a) A DG FeFET   naturally performs $I_{SL} = x\cdot G\cdot y\cdot z$;
  (b) $V_{BG}$ can adjust the $V_{TH}$, thus correspondingly tune $I_{SL}$ with DG FeFET storing '1';
  (c)  $I_{SL}-V_{BG}$  of a  DG FeFET can approximate the partial incremental-E form $E_{inc,p}$, where the fractional annealing factor $f(T)$ is mapped as  $I_{SL}$; 
  (d) The DG FeFET-based crossbar array can implement incremental-E form $E_{inc}$.
  }
  \label{fig: DG FeFET crossbar}
  \vspace{-4ex}
\end{figure}

% Fig. \ref{fig: DG FeFET crossbar}(a) depicts the cell schematic of the DG FeFET-based CiM crossbar array.
% The DG FeFET has four input/output terminals, including DL, FG, BG and SL. 
As shown in Fig. \ref{fig: DG FeFET crossbar}(a), a single DG FeFET has four terminals: the data line (DL), front gate (FG), back gate (BG) and source line (SL). 
A low/high $V_{TH}$ is stored in the DG FeFET, denoted as G = '1'/'0'. 
The inputs at FG, DL and BG are denoted as $x$, $y$ and $z$, where $x$ and $y$ are binary inputs ($x/y \in \{0,1\}$) and $z$ is an analog input. 
SL is the output terminal. 
%The DG FeFET  naturally performs the operation $I_{SL} = x\cdot G\cdot y\cdot z$. 
%can naturally perform the operation $I_{SL} = x\cdot G\cdot y\cdot z$ where $z$ is an analog input applied to BG. 
%To examine the impact of $z$ on the output, we 
%fix the inputs at DL and FG, 
%sweep $V_{BG}$ with fixed DL and FG inputs, and measure the $I_{SL}-V_{BG}$ characteristics, as illustrated in 
Fig. \ref{fig: DG FeFET crossbar}(b) shows the $I_{SL}-V_{BG}$ characteristics given fixed DL and FG inputs. 
It can be seen that DG FeFET  naturally performs the operation $I_{SL} = x\cdot G\cdot y\cdot z$, where $I_{SL}$ remains 0 when G = '0', and follows $V_{BG}$ 
when storing '1'. 
This intrinsic operation of the  DG FeFET resembles  the partial incremental-E formulation, i.e., $E_{inc,p} = \sigma_r \cdot G \cdot \sigma_c \cdot f(T)$. 
% Therefore, to implement incremental-E formulation (Eq. \eqref{equ: incremental-E}), 
%The intrinsic operation of the  DG FeFET, i.e., $I_{SL} = x\cdot G\cdot y\cdot z$, resembles  the partial incremental-E formulation, i.e., $E_{inc,p} = \sigma_r \cdot G \cdot \sigma_c \cdot f(T)$. 
As illustrated in Fig. \ref{fig: DG FeFET crossbar}(c), the normalized output current $I_{SL}$ (with '1' stored) can approximate $E_{inc,p}$ assuming $f(T) \approx \frac{1}{-0.006T+5}-0.2$, 
%Note that $f(T)$ satisfies the constraints outlined in Sec. \ref{subsec:transformation}. 
%Therefore, a single DG FeFET is well-suited for computing $E_{inc,p}$, 
where $G$ is stored as binary  $V_{TH}$ state, $\sigma_r$, $\sigma_c$  are applied to  FG, DL, respectively, and $T$ corresponds to  $V_{BG}$.
$I_{SL}$  is sensed as the value of $E_{inc,p}$. 
% Fig. \ref{fig: DG FeFET crossbar}(c) shows that 
% $I_{SL}$ can approximate the partial energy increment $E_{inc,p}$ by reflecting $x$, $y$ and $z$ to $\sigma_r$, $\sigma_c$ and $f(T)$, respectively. 
% After normalization, the normalized $I_{SL}$

% by mapping $x$, $y$ and $z$ to $\sigma_r$, $\sigma_c$ and $f(T)$, respectively, where $f(T) \approx \frac{10}{-0.006T+5}-2$, satisfying the two constraints in Sec. \ref{subsec:transformation}.

%Based on DF FeFET devices, the DG FeFET-based CiM crossbar array is demonstrated to implement $E_{inc}$, as shown in 
Fig. \ref{fig: DG FeFET crossbar}(d) shows the schematic of the proposed DG FeFET-based crossbar array. 
% Each element in the matrix $J$, e.g., $J_{11}$, is mapped onto a $1\times k$ subarray, %assuming $k$-bit quantization. 
% %Each cell stores 1 bit of the element. 
% with each cell storing 1 bit under $k$-bit quantization. 
% Thus, an $n \times n$ matrix $J$ is mapped onto an $n \times m$ DG FeFET-based crossbar ($m = n\times k$). 
Each row and column share the same FG/BG and DL/SL, respectively. 
The peripheral circuits  include controllers, encoders and drivers for FG, DL and BG inputs, and 
%Specifically, a global control module of BG is required because 
%The output peripherals consist of 
multiplexers (MUXs), ADCs, shift\&adders (S\&As), summations (Sums) and an output buffer for SL outputs. 
Every $k$ columns share the same MUX, ADC, S\&A and Sum. 
Each element in the matrix $J$, e.g., $J_{11}$, is mapped onto a $1\times k$ subarray, %assuming $k$-bit quantization. 
%Each cell stores 1 bit of the element. 
with each cell storing 1 bit under $k$-bit quantization. 
In this way, an $n \times n$ matrix $J$ is mapped onto an $n \times m$  crossbar ($m = n\times k$). 

During each $E_{inc}$ computation, 
the input vectors $\bm{\sigma_r^T}$ and $\bm{\sigma_c}$ are simultaneously directed to FG and DL encoders, respectively, 
and the input scalar $f(T)$ is sent to the BG encoder. 
%The encoders of $\mathbf{\sigma_r^T}$ and $\mathbf{\sigma_c}$, i.e., FG and DL encoders, 
% FG and DL encoders convert the $\bm{\sigma_r^T}$ and $\bm{\sigma_c}$ values into binary voltages and drive the array through  FG and DL drivers, respectively. 
FG and DL encoders convert $\bm{\sigma_r^T}$ and $\bm{\sigma_c}$ into binary voltages and drive the array through  FG and DL drivers, respectively. 
The BG encoder looks at $T$ value and generates the corresponding analog voltage 
%processes the analog value of $f(T)$, generating corresponding analog voltages
via the BG driver to control all BGs. 
%During each incremental-E computation, 
% FGs and DLs receive the corresponding binary bits of input vectors $\mathbf{\sigma_r^T}$ and $\mathbf{\sigma_c}$ from FG and DL drivers, respectively. 
% The BGs are applied with corresponding analog voltage from the BG driver. 
% The currents from SLs are accumulated together to yield the incremental-E value. 
For the array output, $k$ SLs form a group, %and  groups executes 
%with the operations across different groups executed in parallel. 
% Currents from every group of $k$ SLs are sensed sequentially, while the operations across different groups are executed in parallel.
and all groups perform the computations in parallel.  
$k$ output currents from a group are sensed sequentially, multiplexed by  
%Each current is selected by 
the MUX and converted into $k$ digital values via the ADC. 
The $k$ digital values are then shifted, added and summed to generate the partial result. 
These partial results from all groups are  aggregated to compute the final  $E_{inc}$ value,  stored in the output buffer. 
Since the crossbar  only supports non-negative inputs, 
the $E_{inc}$ components associated with positive and negative inputs are  separately calculated, respectively, and summed ultimately.
%its positive and negative components and then summing them. 
%, which will be introduced in Sec. \ref{subsec: flow}
To this end, the DG FeFET-based crossbar array  successfully implements $E_{inc} = \bm{\sigma_{r}^T}  J  \bm{\sigma_c} \cdot f(T)$.

\vspace{-1ex}
\subsection{Tunable BG-based In-Situ Annealing Flow}
\label{subsec: flow}
\vspace{-1ex}

\begin{figure}[!t]
%\vspace{-3ex}
  \centering
  \includegraphics[width=1\columnwidth]{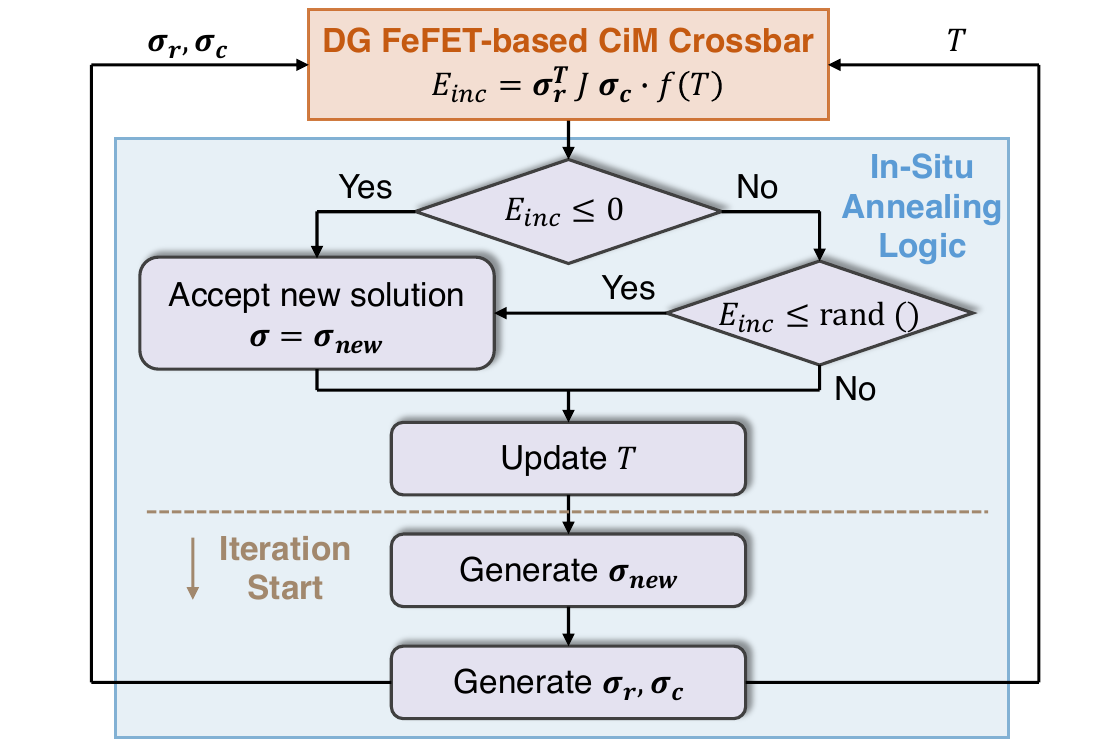}
  \vspace{-5ex}
  \caption{
  Working flow of the proposed CiM   annealer.
  %, including the DG FeFET-based CiM crossbar array and CiM-based simulated annealing (CSA) logic.
  }
  \label{fig: CSA flow}
  \vspace{-4ex}
\end{figure}

Based on the incremental-E transformation and DG FeFET-based CiM crossbar array, we present the proposed CiM  annealer that 
%in-situ annealing algorithm that 
leverages the $V_{TH}$ tunable BG of DG FeFETs for efficient in-situ annealing process.
%and   device-algorithm co-design is employed   
%simulated annealing (CSA) algorithm is proposed 
%for efficient CiM in-situ annealer. 
Fig. \ref{fig: CSA flow} illustrates the work flow of the  annealer. 
In each annealing iteration, a new solution $\bm{\sigma_{new}}$ is randomly generated based on the current solution $\bm{\sigma}$. 
The vectors $\bm{\sigma_r}$ and $\bm{\sigma_c}$ are derived from $\bm{\sigma_{new}}$. %, and 
%the annealing factor $f(T)$ is updated according to the decayed temperature $T$. 
% temperature $T$ is updated. 
The $\bm{\sigma_r}$, $\bm{\sigma_c}$ and temperature $T$ are then directed to the DG FeFET-based CiM crossbar for  in-situ $E_{inc}$ computation. 
%Note that predefined $V_{BG}$ value corresponding to the temperature $T$ is applied to the BGs of DG FeFETs within the crossbar, mapping $f(T)$ directly as the cell current $I_{SL}$, as shown in Fig. \ref{fig: DG FeFET crossbar}(c).
The computed result of $E_{inc}$ is sent to the annealing logic. 
If $E_{inc}\leq 0$, it indicates that the new solution corresponds to lower energy compared to  current solution, 
thus the new solution is accepted, updating the current solution  $\bm{\sigma}=\bm{\sigma_{new}}$. 
If $E_{inc}> 0$, the value is compared with a randomly generated number 
$r\in [0,1]$. 
% $r$. 
%is generated and compared with $E_{inc}$. 
If $E_{inc}\leq r$, the new solution is still accepted; 
otherwise, $\bm{\sigma}$ remains unchanged. 
%The solution 
The current solution $\bm{\sigma}$ then serves as the input of next iteration and  the temperature $T$ is updated. 
The optimal solution is identified through iterative computations. 
Alg. \ref{alg: CSA} presents the pseudocode for the annealing algorithm.

% \setlength{\textfloatsep}{0pt}
% \begin{figure}[!t]
\begin{algorithm}[!t]
  \caption{In-Situ Annealing  Algorithm}
  \label{alg: CSA}
  \begin{algorithmic}[1]
  \Require 
  $J$, $t$ \Comment{Ising model and number of flipped spins ($t=|F|$)}
  % \\ $k$ \Comment{Number of flipped spins in each iteration}
  \Ensure 
  $\bm{\sigma}$  \Comment{Spin configuration that minimizes $E=\bm{\sigma}^TJ\bm{\sigma}$} 
  
  \State Initialize $\bm{\sigma}$ with random values from $\{-1, 1\}$

  \For {each annealing iteration}
      % \State $\bm{\sigma_f}$ = zeros(1, len($J$)) 
      % \State Set $k$ random elements of $\bm{\sigma_f}$ to 1 \Comment{Marks spins to flip}
      \State Select $t$ elements and get $\bm{\sigma_f}$ \Comment{Marks spins to flip}
      \State $\bm{\sigma_{new}} = \bm{\sigma} \circ (1 - 2 \cdot \bm{\sigma_f})$ \Comment{Flip elements in $\bm{\sigma}$}
    
      \State $\bm{\sigma_c} = \bm{\sigma_{new}} \circ \bm{\sigma_f}$, $\bm{\sigma_r} = \bm{\sigma_{new}} \circ (1 - \bm{\sigma_f})$ %\Comment{Flipped elements in $\bm{\sigma_{new}}$}
      % \State $\bm{\sigma_r} = \bm{\sigma_{new}} \circ (1 - \bm{\sigma_f})$ %\Comment{Unflipped elements in $\bm{\sigma_{new}}$}
    
      \State $E_{inc} = \bm{\sigma_r}^T  J  \bm{\sigma_c} \cdot f(T)$ \Comment{DG FeFET-based CiM acceleration}
      \If {$E_{inc} \leq 0$}
        \State $\bm{\sigma} = \bm{\sigma_{new}}$ \Comment{Accept new solution}
      \Else
        % \State $r =$ random(0, 1)
        \If {$E_{inc} \leq random(0, 1)$}
        \State $\bm{\sigma} = \bm{\sigma_{new}}$ \Comment{Accept new solution}
        \EndIf
      \EndIf
      \State Update $T$ based on annealing schedule
  \EndFor
% \vspace{-1ex}
\end{algorithmic}
\end{algorithm}
% \vspace*{-1ex}
% \setlength{\textfloatsep}{20pt}
% \end{figure}

%During the annealing, the temperature $T$ decreases from  high to  low, 
% Since $T$ is transformed into $f(T)$ and applied as a voltage to the BGs of the DG FeFET-based crossbar, the temperature variation range needs to be mapped to the voltage range of $f(T)$. 
The range of annealing temperature $T$ is normalized to match the BG voltage range $V_{BG}$ (0.7V to 0V) for the DG FeFET-based crossbar with a gradient of 0.01V, such that the temperature related annealing factor $f(T)$ can be approximately mapped to the crossbar currents by tuning $V_{BG}$, as illustrated in Fig. \ref{fig: DG FeFET crossbar}(c).
% Specifically, $f(T)$ is normalized to a range of 0.7V to 0V to match the voltage input limitations of the circuit, with a gradient set at 0.01V.
$T$ does not necessarily decrease with each iteration. 
%In cases where a large number of iterations are required, the temperature may only be reduced after a certain number of iterations. 
For longer processes, $T$ decreases only after a pre-set number of iterations.
Once $V_{BG}$ reaches 0V, it remains at zero, terminating the annealing.

%% file: 04_Evaluation.tex
%\vspace{-1ex}
\section{Evaluation}
\label{sec: evaluation}
\vspace{-1ex}

In this section, we first compare the hardware overhead of the proposed CiM annealer  to other two CiM annealers \cite{hussain2022area, yin2024ferroelectric}. 
We then evaluate  their efficiency in solving COPs. 
%compared to alternative approaches.
All simulations were conducted using SPECTRE. 
%The 22nm BSIM IMG based DG FeFET model \cite{jiang2022asymmetric} is adopted for DG FeFET, 
The 22nm BSIM IMG DG FeFET model \cite{jiang2022asymmetric} and Preisach FeFET model \cite{ni2018circuit} are adopted for DG FeFET and FeFET, respectively. 
A 22nm commercial model is used for MOSFETs with TT process corner at a temperature of 27$^{\circ}$C. 
Each annealer contains a single crossbar. 
%Every eight columns within the crossbar share an
8-to-1 multiplexed ADCs are employed \cite{liu202016}, which are scaled to 22nm.
%and corresponding precision. 
% Each annealer contains a single crossbar, with every eight columns sharing an ADC. 
% The ADC is adopted from \cite{liu202016} and scaled to 22nm with appropriate precision. 
The hardware overhead for  the exponential function implementation is derived from \cite{hussain2022area}. 
% which is scaled from a 22nm SAR ADC model \cite{liu202016} with appropriate precision.
% The SAR ADC \cite{liu202016} is scaled to 22nm and 10-12 bit to measure the output currents of crossbar.
The wiring parasitics are extracted from DESTINY \cite{poremba2015destiny}. 
The Max-Cut problem is selected as the representative COP \cite{maxcutstanford}.

% In this section, we first validate the functionality of our proposed FeFET-based inequality filter and evaluate the hardware overhead reduction achieved by HyCiM in comparison to D-QUBO based annealer. 
% We then evaluate the efficiency improvements in solving COPs with HyCiM compared to others.
% %D-QUBO based annealer.
% All simulations were conducted using SPECTRE.
% %and the FeFET model employed is the Preisach FeFET model \cite{ni2018A}.
% The Preisach FeFET model \cite{ni2018A} is adopted for FeFET, the 28nm TSMC model is used for MOSFETs with TT process corner at a temperature of 27$^{\circ}$C.
% %To model the MOSFETs, we utilize the 40nm UMC model \cite{vattikonda2006modeling} with TT process corner at a temperature of 27$^{\circ}$C. 
% The wiring parasitics
% %for the 40nm technology node 
% are extracted from DESTINY \cite{poremba2015destiny}. 
% %For our evaluation, 
% We choose QKP as the representative COP and select 40 instances from \cite{QKPdataset}, each containing 100 items.
% %. Each problem instance consists of 100 items.

\vspace{-1ex}
\subsection{Hardware Overhead}
\vspace{-1ex}
\label{subsec: hardware overhead}

\begin{figure}[!t]
%\vspace{-3ex}
  \centering
  \includegraphics[width=1\columnwidth]{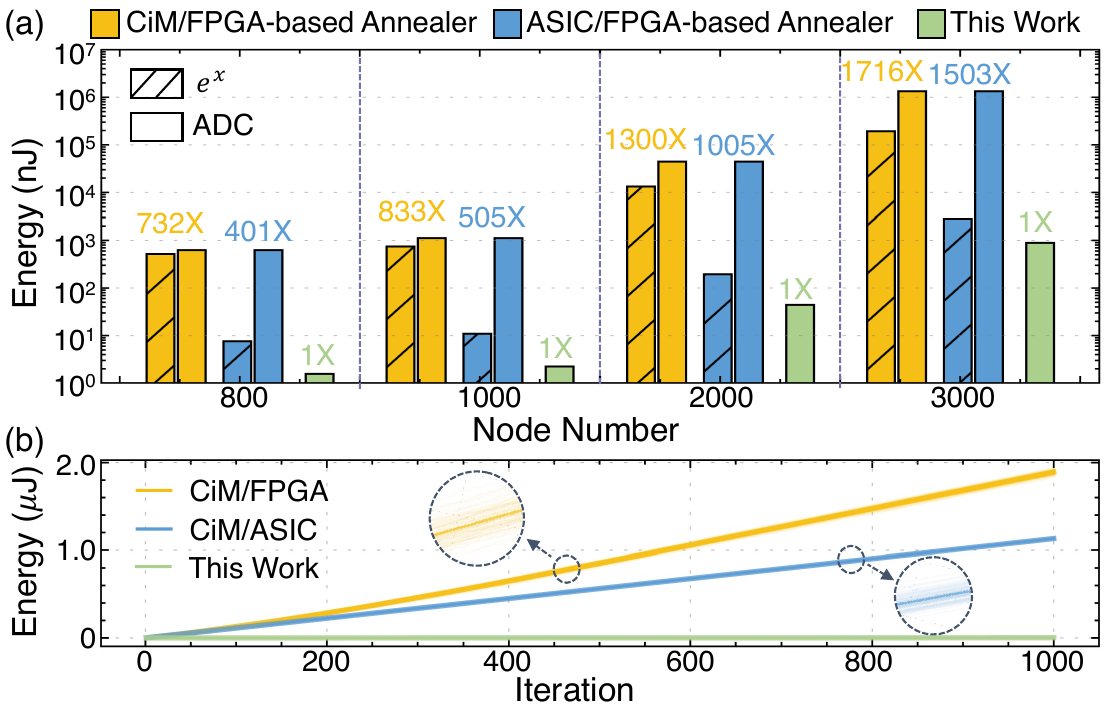}
  \vspace{-6ex}
  \caption{
  %Energy  of different CiM annealers. 
  (a) Average energy  in solving 30 Max-Cut problem instances and
  (b) energy trends in solving a 1000-node instance across three CiM annealers.
  }
  \label{fig: energy}
  \vspace{-4ex}
\end{figure}

\begin{figure}[!t]
%\vspace{-3ex}
  \centering
  \includegraphics[width=1\columnwidth]{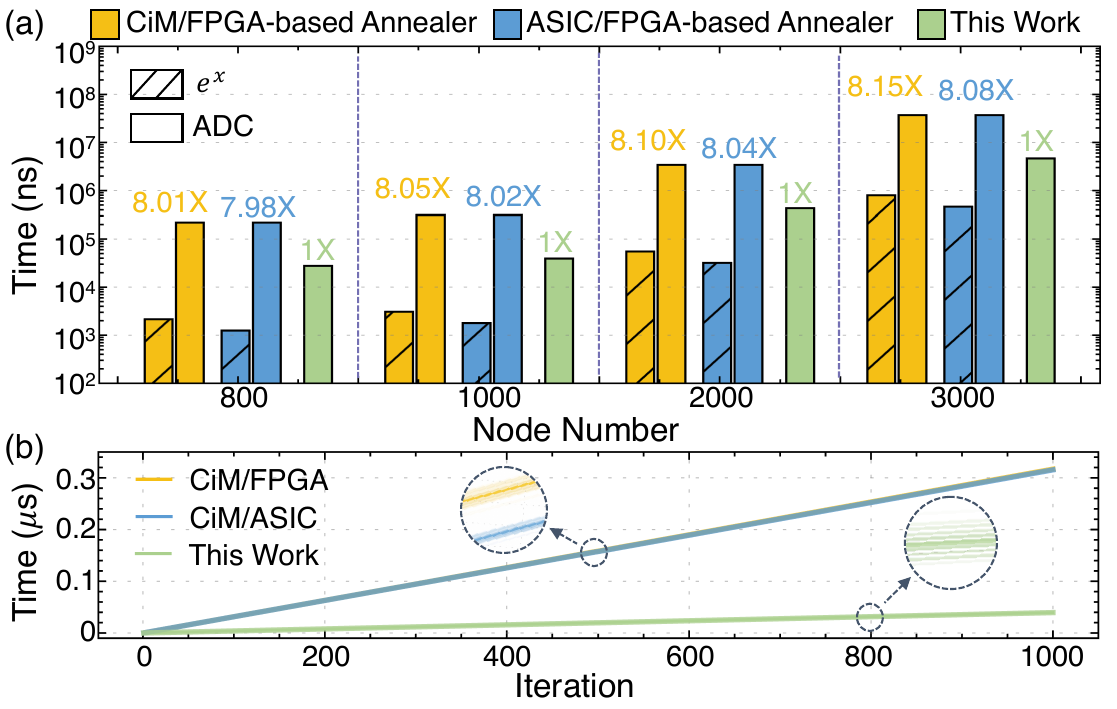}
  \vspace{-6ex}
  \caption{
  %Time cost of different CiM annealers. 
  (a) Average time cost in solving 30 Max-Cut problems and
  (b)  time cost trends in solving a 1000-node instance across three CiM annealers.
  }
  \label{fig: time}
  \vspace{-4ex}
\end{figure}

% In our evaluation, the crossbar arrays are of $n\times n$ size, where $n$ is the node number of the Max-Cut problem. 
% We compare the hardware overhead of our proposed in-situ DG FeFET-based CiM annealer with the state-of-art FeFET-based CiM annealers \cite{yin2024ferroelectric}. 
% To implement the exponential function, the FPGA and ASIC implementations are adopted \cite{hussain2022area}. 
% In our evaluation, we compare the hardware overhead of our proposed in-situ DG FeFET-based CiM annealer with two other FeFET-based CiM annealers. 
For a fair comparison, the other two annealers are based on FeFET CiM accelerations \cite{yin2024ferroelectric} and FPGA/ASIC modules for exponential computation \cite{hussain2022area}, denoted as CiM/FPGA and CiM/ASIC-based annealers, respectively. 
% Both annealers implement the conventional SA algorithm, utilizing FeFET crossbars for the VMV function \cite{yin2024ferroelectric} and FPGA/ASIC implementations for the exponential function \cite{hussain2022area}. 
% For simplicity, we refer to these as the FPGA-based and ASIC-based annealers, respectively. 
%We randomly selected four Max-Cut instances with varying node counts: Gset1 (800 nodes), Gset43 (1000 nodes), Gset22 (2000 nodes), and Gset48 (3000 nodes). 
We selected 30 Max-Cut problem instances across four groups: 9 with 800 nodes, 9 with 1000 nodes, 9 with 2000 nodes, and 3 with 3000 nodes. 
For each instance, we conducted 100 runs. 
The iteration numbers per run are set according to problem size, with 700/1000/10,000/100,000 iterations  for the 800-/1000-/2000-/3000-node instances, respectively.

Fig. \ref{fig: energy}(a) illustrates the average energy consumption of the three annealers. 
The major energy consumption are from the ADC and the exponential function implementation.
%, while the crossbar's energy is omitted as it is negligible. 
It can be seen that, 
our proposed in-situ annealer achieves significant energy savings by reducing the ADC execution times and eliminating the energy  associated with $e^x$ calculation. 
This efficiency stems from the incremental-E transformation, which removes redundant VMV multiplications  and simplifies the exponential calculation to a fractional function.
%, thus reducing the computational complexity.
Such  transformation fully takes advantages of the DG FeFET-based CiM array,  enabling efficient in-situ $E_{inc}$ computations. 
Notably, the  energy efficiency improvement increases as the problem size grows. 
% This is because larger COPs requires more annealing iterations, enlarging the energy saving the reduction in ADC energy consumption is more substantial, which highlights the potential of the proposed annealer for efficiently solving large-scale COPs. 
Fig. \ref{fig: energy}(b) shows the energy consumptions versus iteration count of the three annealers when solving a 1000-node instance. 
The CiM/FPGA and CiM/ASIC-based annealers display a rapid linear energy increase with iteration counts.
%show negligible variations in energy consumption across different runs 
% Because both implementations  execute    $e^x$ computation based on the sign of $\Delta E$ within each iteration, negligible energy variations across different are observed.
%the overall energy consumption trends display a rapid linear increase with iteration counts. 
Conversely, our proposed  annealer 
%reduces the overall computational complexity and eliminates the costly $e^x$ computation, thus  
demonstrates a significantly slower rise in energy consumption. 
%indicating its superior efficiency for larger problem sizes.
This is because larger-scale COPs require more annealing iterations, enlarging the energy savings associated with reduced ADC executions and simplified fractional  factor calculations.
%, which highlights the superiority of out approach for  solving large-scale COPs. 

Fig. \ref{fig: time}(a) presents the average  time  for problem solving. 
The proposed annealer shows substantial time savings across different problem sizes. 
% This efficiency is achieved by reducing ADC operations. 
Unlike prior annealers that activate  the entire  array for energy computation, the proposed annealer only activates specific columns associated with updated spins, thus significantly reducing the overall ADC time cost. 
Fig. \ref{fig: time}(b) shows the time cost trends in solving a 1000-node instance. 
The other two annealers have  identical trends due to the dominant ADC time. 
%$Unlike the steady energy evolution of the proposed annealer in Fig. \ref{fig: energy}(b), 
The proposed annealer shows much less time, as it eliminates significant ADC executions associated with redundant energy form computations. 
%This is because a single ADC may need to measure currents from multiple columns sequentially, as each ADC is shared among eight columns. 

\begin{figure}[!t]
%\vspace{-3ex}
  \centering
  \includegraphics[width=1\columnwidth]{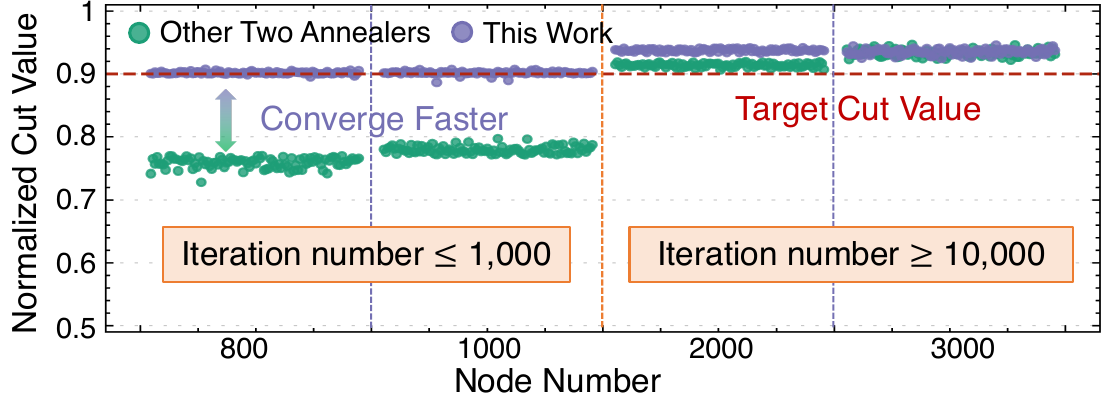}
  \vspace{-6ex}
  \caption{
  Normalized Cut values solved by the proposed  CiM in-situ annealer and other annealers.
  }
  \label{fig: accuracy}
  \vspace{-3.5ex}
\end{figure}

\begin{table}\huge%[!t]\large%\vspace{-0.3cm}
\caption{Summary of COP Solvers}
\vspace{-2ex}
\label{table:evaluation}
\centering
\resizebox{\columnwidth}{!}{
\begin{tabular}{|c|c|c|c|c|c|c|}
\toprule
\hline
Reference & \cite{cai2020power}& \cite{yin2024ferroelectric}& \cite{taoka2021simulated}& \cite{qian2024hycim} & \cite{qian2024c} & This Work       \\ \hline

\multirow{2}{*}{COP}           & \multirow{2}{*}{Max-Cut} & Graph & \multirow{2}{*}{Knapsack} & Quadratic & Nash  & \multirow{2}{*}{Max-Cut} \\ 
                              & & Coloring  &  & Knapsack & Equilibrium & \\ \hline
                              
Complexity                    & $O(n^2)$  & $O(n^2)$ & $O(n^2)$ & $O(n^2)$ & $O(n^2)$  & $O(n)$  \\ \hline

$e^x$ Computation & Yes  & Yes & Yes & Yes   & Yes & No   \\ \hline

Hardware Imp.$^\star$ & \multirow{2}{*}{Memristor} &  \multirow{2}{*}{FeFET}   & \multirow{2}{*}{RRAM}    & \multirow{2}{*}{FeFET}    & \multirow{2}{*}{FeFET}     & DG  \\ 
                            for Crossbar  & &  &  &  &   &FeFET  \\ \hline
                            
Problem Size                  & 60 node  & 21 node & 10 node & 100 node & 104 node  & 3000 node \\ \hline

Time to & \multirow{2}{*}{6.6$\mu$s} &  \multirow{2}{*}{5.1$\mu$s$^\ddagger$}   & \multirow{2}{*}{3.8$\mu$s$^\ddagger$}    & \multirow{2}{*}{1.3ms$^\ddagger$}    & \multirow{2}{*}{0.08s}     & \multirow{2}{*}{4.6ms}  \\ 
Solution  & &  &  &  &   &  \\ \hline

Energy to & \multirow{2}{*}{0.07$\mu$J} &  \multirow{2}{*}{0.2$\mu$J$^\ddagger$}   & \multirow{2}{*}{-}    & \multirow{2}{*}{2.1$\mu$J$^\ddagger$}    & \multirow{2}{*}{-}     & \multirow{2}{*}{0.9$\mu$J}  \\ 
Solution  & &  &  &  &   &  \\ \hline

Average      & \multirow{2}{*}{65$^\dagger$}        & \multirow{2}{*}{-}        & \multirow{2}{*}{92.4$^\dagger$}     & \multirow{2}{*}{98.54}     & \multirow{2}{*}{81.9}      & \multirow{2}{*}{98}         \\ 
                            Success Rate (\%)  & & &  &  &   &  \\ \hline
\bottomrule
\end{tabular}
}
\vspace{-0.5ex}
\begin{flushleft}
\scriptsize
$^\star$: Imp.: implementation.
$^\dagger$: Extracted from literature.
$^\ddagger$: Extracted from literature and estimation at  22nm  node. 
\end{flushleft}
\vspace{-2.5ex}
\end{table}

\vspace{-1ex}
\subsection{Problem Solving Efficiency}
\vspace{-1ex}
\label{subsec: efficiency}

We  compare the COP solving efficiency of the %proposed in-situ CiM annealer compared to FPGA/ASIC-based
three CiM annealers. 
Because CiM/FPGA and CiM/ASIC-based annealers adopt the same algorithm, the results are consistent. 
% We selected 30 Max-Cut problems across four groups: 9 with 800 nodes (Gset1-5, 14-17), 9 with 1000 nodes (Gset43-47, 51-54), 9 with 2000 nodes (Gset22-26, 35-38), and 3 with 3000 nodes (Gset48-50). 
% The iteration numbers for each group follow the same setup in Sec. \ref{subsec: hardware overhead}. 
% For each instance, 100 runs were performed on the proposed and current annealers. 
We used Monte Carlo sampling to select 100 runs for each Max-Cut problem group, and the normalized Cut values are shown in Fig. \ref{fig: accuracy}. 
The target Cut value is set as 90\% of the true optimal value, suggesting a successful instance solving. 
The results clearly demonstrate that the proposed annealer 
consistently approaches the optimal solution for all COPs, achieving an impressive average success rate of 98\%. 
In contrast, the other two annealers show a lower average success rate of 50\%, only solving  the 2000-node and 3000-node problems successfully. 
%This suggests that although all annealers can solve the problems ultimately, the proposed annealer converges faster given limited iteration counts, demonstrating its higher solving efficiency. 
Interesting, the fractional annealing factor approximately implemented by DG FeFET's normalized current enables faster convergence than other annealers with exponential annealing factor, achieving higher solving efficiency.
%counts exponential annealing factor of other annealer is replaced by fractional function, and then approximated by the DG FeFET current, 
%\polish{
%This is because the proposed annealer takes advantage of the inherent characteristics of DG FeFET for tunable BG-based in-situ annealing, thus improving the efficiency. 
%}
The solver summary in Table. \ref{table:evaluation} shows that the proposed  CiM in-situ annealer can solve larger COPs with high energy and time efficiency,  achieving a high success rate.

%% file: 05_Conclusion.tex
\vspace{-0ex}
\section{Conclusion}
\label{sec:conclusion}
\vspace{-1ex}
%In this paper, we propose a general compact and energy efficient CAM design that employs the minimum number of NVM as storage element to alleviate the design overhead and optimize the search energy. 
%We propose a novel 2T-1FeFET CAM design which utilizes just one FeFET and can be adoped to other NVMs. We then propose an adaptive $ML$ precharge and discharge scheme implemented by a TIQ comparator based SA for further energy optimization. Evaluation results and  query processing application benchmarking suggest that our proposed 2T-1FeFET CAM array with $ML$ optimization scheme achieves better energy efficiency and performance when compared with other state-of-the-art CAM approaches.

%In this paper, we propose a hybrid CiM QUBO solver architecture named HyCiM to accelerate solving COPs with inequality constraints.
%We propose a novel transformation method to transform a COP with inequality constraints into inequality-QUBO form. We then propose a FeFET-based CiM inequality filter to filter out feasible solutions and reduce search space and hardware overhead. A FeFET-based crossbar is proposed for QUBO acceleration, and SA mechanism is employed to improve solving effiency.
%Evaluation results show that compared to other D-QUBO based solvers, HyCiM has significant improvement in search space, hardware overhead and problem solving efficiency.

In this paper, we introduce a ferroelectric CiM in-situ annealer  to efficiently solve COPs. 
We present an incremental-E transformation method that converts a COP into an $E_{inc}$ form, reducing computational complexity and eliminating costly exponential function calculations. 
We propose a DG FeFET-based CiM crossbar to implement the $E_{inc}$ form by utilizing the unique characteristic and structure of DG FeFETs. 
Overall, we build a device-algorithm co-design framework that accommodates efficient in-situ $E_{inc}$  computation and annealing flow to solve COPs. 
Evaluation results show that the proposed annealer significantly reduces energy consumption and time costs while improving the problem solving efficiency compared to prior counterpart solvers.

% In this paper, we introduce HyCiM, a hybrid CiM QUBO solver framework designed to efficiently solve COPs with inequality constraints. 
% We present a novel transformation method that converts COPs with inequality constraints into an inequality-QUBO form.
% We propose a FeFET based inequality filter that evaluates inequalities and filters out infeasible inputs by leveraging the multi-level characteristic of FeFETs.
% %, thus reducing the search space and hardware overhead. 
% We fabricate a FeFET based CiM crossbar that accelerates the QUBO computations by exploiting the 
% %, which is implemented by our proposed FeFET-based CiM inequality filter and crossbar array. 
% single transistor multiplication property of FeFET devices as well as SA process.
% %are leveraged in inequality filter and crossbar, respectively, to filter feasible input configurations and accelerate QUBO form, effectively reducing the search space and hardware overhead. The proposed FeFET-based crossbar accelerates QUBO computations, complemented by the use of SA mechanism to enhance solving efficiency.
% Evaluation results show that HyCiM offers substantial search space reduction, hardware overhead saving, and improved problem solving efficiency compared to prior QUBO solvers.
\vspace{-1ex}

%% file: 00_Main.bbl
\begin{thebibliography}{10}
\providecommand{\url}[1]{#1}
\csname url@samestyle\endcsname
\providecommand{\newblock}{\relax}
\providecommand{\bibinfo}[2]{#2}
\providecommand{\BIBentrySTDinterwordspacing}{\spaceskip=0pt\relax}
\providecommand{\BIBentryALTinterwordstretchfactor}{4}
\providecommand{\BIBentryALTinterwordspacing}{\spaceskip=\fontdimen2\font plus
\BIBentryALTinterwordstretchfactor\fontdimen3\font minus \fontdimen4\font\relax}
\providecommand{\BIBforeignlanguage}[2]{{%
\expandafter\ifx\csname l@#1\endcsname\relax
\typeout{** WARNING: IEEEtran.bst: No hyphenation pattern has been}%
\typeout{** loaded for the language `#1'. Using the pattern for}%
\typeout{** the default language instead.}%
\else
\language=\csname l@#1\endcsname
\fi
#2}}
\providecommand{\BIBdecl}{\relax}
\BIBdecl

\bibitem{yu2013industrial}
G.~Yu, \emph{Industrial applications of combinatorial optimization}.\hskip 1em plus 0.5em minus 0.4em\relax Springer Science \& Business Media, 2013, vol.~16.

\bibitem{paschos2014applications}
V.~T. Paschos, \emph{Applications of combinatorial optimization}.\hskip 1em plus 0.5em minus 0.4em\relax John Wiley \& Sons, 2014, vol.~3.

\bibitem{naseri2020application}
G.~Naseri and M.~A. Koffas, ``Application of combinatorial optimization strategies in synthetic biology,'' \emph{Nature communications}, vol.~11, no.~1, p. 2446, 2020.

\bibitem{barahona1988application}
F.~Barahona, M.~Gr{\"o}tschel, M.~J{\"u}nger \emph{et~al.}, ``An application of combinatorial optimization to statistical physics and circuit layout design,'' \emph{Operations Research}, vol.~36, no.~3, pp. 493--513, 1988.

\bibitem{jiang2023efficient}
M.~Jiang, K.~Shan, C.~He \emph{et~al.}, ``Efficient combinatorial optimization by quantum-inspired parallel annealing in analogue memristor crossbar,'' \emph{Nature Communications}, vol.~14, no.~1, p. 5927, 2023.

\bibitem{yue2024scalable}
W.~Yue, T.~Zhang, Z.~Jing \emph{et~al.}, ``A scalable universal ising machine based on interaction-centric storage and compute-in-memory,'' \emph{Nature Electronics}, vol.~7, no.~10, pp. 904--913, 2024.

\bibitem{yin2024ferroelectric}
X.~Yin, Y.~Qian, A.~Vardar \emph{et~al.}, ``Ferroelectric compute-in-memory annealer for combinatorial optimization problems,'' \emph{Nature Communications}, vol.~15, no.~1, p. 2419, 2024.

\bibitem{mohseni2022ising}
N.~Mohseni, P.~L. McMahon, and T.~Byrnes, ``Ising machines as hardware solvers of combinatorial optimization problems,'' \emph{Nature Reviews Physics}, vol.~4, pp. 363--379, 2022.

\bibitem{katsuki2022fast}
K.~Katsuki, D.~Shin, N.~Onizawa \emph{et~al.}, ``Fast solving complete 2000-node optimization using stochastic-computing simulated annealing,'' in \emph{2022 29th IEEE International Conference on Electronics, Circuits and Systems (ICECS)}.\hskip 1em plus 0.5em minus 0.4em\relax IEEE, 2022, pp. 1--4.

\bibitem{takemoto20214}
T.~Takemoto, K.~Yamamoto, C.~Yoshimura \emph{et~al.}, ``4.6 a 144kb annealing system composed of 9$\times$ 16kb annealing processor chips with scalable chip-to-chip connections for large-scale combinatorial optimization problems,'' in \emph{2021 IEEE International Solid-State Circuits Conference (ISSCC)}, vol.~64.\hskip 1em plus 0.5em minus 0.4em\relax IEEE, 2021, pp. 64--66.

\bibitem{moy20221}
W.~Moy, I.~Ahmed, P.-w. Chiu \emph{et~al.}, ``A 1,968-node coupled ring oscillator circuit for combinatorial optimization problem solving,'' \emph{Nature Electronics}, vol.~5, no.~5, pp. 310--317, 2022.

\bibitem{ahmed2021probabilistic}
I.~Ahmed, P.-W. Chiu, W.~Moy \emph{et~al.}, ``A probabilistic compute fabric based on coupled ring oscillators for solving combinatorial optimization problems,'' \emph{IEEE Journal of Solid-State Circuits}, vol.~56, no.~9, pp. 2870--2880, 2021.

\bibitem{taoka2021simulated}
K.~Taoka, N.~Misawa, S.~Koshino \emph{et~al.}, ``Simulated annealing algorithm \& reram device co-optimization for computation-in-memory,'' in \emph{2021 IEEE International Memory Workshop (IMW)}.\hskip 1em plus 0.5em minus 0.4em\relax IEEE, 2021, pp. 1--4.

\bibitem{qian2024c}
Y.~Qian, K.~Ni, T.~Kampfe \emph{et~al.}, ``C-nash: A novel ferroelectric computing-in-memory architecture for solving mixed strategy nash equilibrium,'' in \emph{Proceedings of the 61st ACM/IEEE Design Automation Conference}, 2024, pp. 1--6.

\bibitem{qian2024hycim}
Y.~Qian, Z.~Yang, K.~Ni \emph{et~al.}, ``Hycim: A hybrid computing-in-memory qubo solver for general combinatorial optimization problems with inequality constraints,'' in \emph{Proceedings of the 61st ACM/IEEE Design Automation Conference}, 2024, pp. 1--6.

\bibitem{qian2025ferroelectric}
Y.~Qian, D.~Huang, A.~Vardar \emph{et~al.}, ``Ferroelectric compute-in-memory framework for solving pure and mixed strategy nash equilibrium,'' \emph{IEEE Transactions on Circuits and Systems I: Regular Papers}, 2025.

\bibitem{kim2024scaling}
S.~Kim, S.~Um, W.~Jo \emph{et~al.}, ``Scaling-cim: edram in-memory-computing accelerator with dynamic-scaling adc and adaptive analog operation,'' \emph{IEEE Journal of Solid-State Circuits}, 2024.

\bibitem{hussain2022area}
M.~A. Hussain, S.-W. Lin, and T.-H. Tsai, ``An area-efficient and high throughput hardware implementation of exponent function,'' in \emph{2022 IEEE International Symposium on Circuits and Systems (ISCAS)}.\hskip 1em plus 0.5em minus 0.4em\relax IEEE, 2022, pp. 3369--3372.

\bibitem{glover2018tutorial}
F.~Glover, G.~Kochenberger, and Y.~Du, ``A tutorial on formulating and using qubo models,'' \emph{arXiv preprint arXiv:1811.11538}, 2018.

\bibitem{mallick2023cmos}
A.~Mallick, Z.~Zhao, M.~K. Bashar \emph{et~al.}, ``Cmos-compatible ising machines built using bistable latches coupled through ferroelectric transistor arrays,'' \emph{Scientific reports}, vol.~13, p. 1515, 2023.

\bibitem{afoakwa2021brim}
R.~Afoakwa, Y.~Zhang, U.~K.~R. Vengalam \emph{et~al.}, ``Brim: bistable resistively-coupled ising machine,'' in \emph{2021 IEEE International Symposium on High-Performance Computer Architecture (HPCA)}.\hskip 1em plus 0.5em minus 0.4em\relax IEEE, 2021, pp. 749--760.

\bibitem{zhuo2022design}
C.~Zhuo, Z.~Yang, K.~Ni \emph{et~al.}, ``Design of ultra-compact content addressable memory exploiting 1t-1mtj cell,'' \emph{IEEE Transactions on Computer-Aided Design of Integrated Circuits and Systems}, 2022.

\bibitem{qian2024enhancing}
Y.~Qian, L.~Zhao, F.~Meng \emph{et~al.}, ``Enhancing convnets with convfifo: A crossbar pim architecture based on kernel-stationary first-in-first-out dataflow,'' \emph{IEEE Transactions on Very Large Scale Integration (VLSI) Systems}, no.~01, pp. 1--12, 2024.

\bibitem{zhao2024convfifo}
L.~Zhao, Y.~Qian, F.~Meng \emph{et~al.}, ``Convfifo: A crossbar memory pim architecture for convnets featuring first-in-first-out dataflow,'' in \emph{2024 29th Asia and South Pacific Design Automation Conference (ASP-DAC)}.\hskip 1em plus 0.5em minus 0.4em\relax IEEE, 2024, pp. 824--829.

\bibitem{mondal2018situ}
A.~Mondal and A.~Srivastava, ``In-situ stochastic training of mtj crossbar based neural networks,'' in \emph{Proceedings of the International Symposium on Low Power Electronics and Design}, 2018, pp. 1--6.

\bibitem{yin2024homogeneous}
X.~Yin, Q.~Huang, H.~E. Barkam \emph{et~al.}, ``A homogeneous fefet-based time-domain compute-in-memory fabric for matrix-vector multiplication and associative search,'' \emph{IEEE TCAD}, 2024.

\bibitem{ni2019ferroelectric}
K.~Ni, X.~Yin, A.~F. Laguna \emph{et~al.}, ``Ferroelectric ternary content-addressable memory for one-shot learning,'' \emph{Nature Electronics}, vol.~2, no.~11, pp. 521--529, 2019.

\bibitem{xu2023challenges}
H.~Xu, J.~Yang, T.~K{\"a}mpfe \emph{et~al.}, ``On the challenges and design mitigations of single transistor ferroelectric content addressable memory,'' \emph{IEEE EDL}, 2023.

\bibitem{cai2024scalable}
J.~Cai, H.~E. Barkam, M.~Imani \emph{et~al.}, ``A scalable 2t-1fefet based content addressable memory design for energy efficient data search,'' \emph{IEEE TCAD}, 2024.

\bibitem{yin2023ultracompact}
X.~Yin, F.~M{\"u}ller, Q.~Huang \emph{et~al.}, ``An ultracompact single-ferroelectric field-effect transistor binary and multibit associative search engine,'' \emph{Advanced Intelligent Systems}, vol.~5, no.~7, p. 2200428, 2023.

\bibitem{yin2024deep}
X.~Yin, F.~M{\"u}ller, A.~F. Laguna \emph{et~al.}, ``Deep random forest with ferroelectric analog content addressable memory,'' \emph{Science Advances}, vol.~10, p. eadk8471, 2024.

\bibitem{hu2021memory}
X.~S. Hu, M.~Niemier, A.~Kazemi \emph{et~al.}, ``In-memory computing with associative memories: A cross-layer perspective,'' in \emph{2021 IEEE International Electron Devices Meeting (IEDM)}.\hskip 1em plus 0.5em minus 0.4em\relax IEEE, 2021, pp. 25--2.

\bibitem{yin2022ferroelectric}
X.~Yin, Y.~Qian, M.~Imani \emph{et~al.}, ``Ferroelectric ternary content addressable memories for energy-efficient associative search,'' \emph{IEEE Transactions on Computer-Aided Design of Integrated Circuits and Systems}, vol.~42, no.~4, pp. 1099--1112, 2022.

\bibitem{jiang2022asymmetric}
Z.~Jiang, Y.~Xiao, S.~Chatterjee \emph{et~al.}, ``Asymmetric double-gate ferroelectric fet to decouple the tradeoff between thickness scaling and memory window,'' in \emph{2022 IEEE Symposium on VLSI Technology and Circuits (VLSI Technology and Circuits)}.\hskip 1em plus 0.5em minus 0.4em\relax IEEE, 2022, pp. 395--396.

\bibitem{ni2018circuit}
K.~Ni, M.~Jerry, J.~A. Smith \emph{et~al.}, ``A circuit compatible accurate compact model for ferroelectric-fets,'' in \emph{2018 IEEE symposium on VLSI technology}.\hskip 1em plus 0.5em minus 0.4em\relax IEEE, 2018, pp. 131--132.

\bibitem{liu202016}
J.~Liu, X.~Tang, W.~Zhao \emph{et~al.}, ``16.5 a 13b 0.005 mm 2 40ms/s sar adc with kt/c noise cancellation,'' in \emph{2020 IEEE International Solid-State Circuits Conference-(ISSCC)}.\hskip 1em plus 0.5em minus 0.4em\relax IEEE, 2020, pp. 258--260.

\bibitem{poremba2015destiny}
M.~Poremba, S.~Mittal, D.~Li \emph{et~al.}, ``Destiny: A tool for modeling emerging 3d nvm and edram caches,'' in \emph{2015 Design, Automation \& Test in Europe Conference \& Exhibition (DATE)}.\hskip 1em plus 0.5em minus 0.4em\relax IEEE, 2015, pp. 1543--1546.

\bibitem{maxcutstanford}
``{Stanford Max-Cut dataset},'' \url{https://web.stanford.edu/~yyye/yyye/Gset/}.

\bibitem{cai2020power}
F.~Cai, S.~Kumar, T.~Van~Vaerenbergh \emph{et~al.}, ``Power-efficient combinatorial optimization using intrinsic noise in memristor hopfield neural networks,'' \emph{Nature Electronics}, vol.~3, no.~7, pp. 409--418, 2020.

\end{thebibliography}
